\documentclass[%
 reprint, amsmath,amssymb, prb,floatfix,
]{revtex4-2}

\usepackage{graphicx}% Include figure files
\usepackage{color}
\usepackage[normalem]{ulem}

\def\ket#1{|#1\rangle }
\def\bra#1{\langle#1 | }

\def\punkt{\;\; .}

\def\expect#1{\langle#1 \rangle}

\def\w{\omega}

\def\e{\epsilon}
\def\k{\vec{k}}
\def\non{\nonumber\\ }

\usepackage[nolist]{acronym}		%only print the acronyms, which we have used
\begin{document}

\title{Open Wilson chain  numerical renormalization group approach to Green's functions}

\author{Jan B\"oker }
\author{Frithjof B. Anders}
\affiliation{Condensed Matter Theory, Department of Physik, Technische Universit\"at Dortmund
Otto-Hahn-Str. 4, 44227 Dortmund,
Germany}

\date{\today}

\begin{abstract}
By combining Wilson's \ac{NRG} with a modified \ac{BRA} we are able to
eliminate the artificial broadening of the Lehmann representation of
quantum impurity spectral functions required by the standard \ac{NRG}
algorithm. Our approach is based on the exact reproduction of the
continuous coupling function in the original quantum impurity
model. It augments each chain site of the Wilson chain by a coupling
to an additional reservoir. This open Wilson chain is constructed by a
continuous fraction expansion and the coupling function is treated in
second order in the context of the \ac{BRA}. The eigenvalues of the
resulting \ac{BRT} generate a finite life time of the \ac{NRG}
excitations that leads to a natural broadening of the spectral
functions. We combine this approach with z-averaging and an analytical
exact expression for the correlation part of the self-energy to obtain
an accurate representation of the spectral function of the original
continuum model in the absence and presence of an external magnetic
field.

\end{abstract}
\maketitle

\begin{acronym}[DMRG]		%give here the longest word from the list 
 \acro{GF}{Green's function}
 \acro{NRG}{numerical renormalization group}
 \acro{ED}{exact diagonalization}
 \acro{DMRG}{density matrix renormalization group}
 \acro{QIS}{Quantum impurity system}
 \acro{DOF}{degrees of freedom}
 \acro{STM}{scanning tunneling microscopy}
 \acro{STS}{scanning tunneling spectra}
 \acro{DMFT}{dynamical mean field theory}
 \acro{SBM}{spin boson model}
 \acro{SIAM}{single impurity Anderson model}
 \acro{BRA}{Bloch-Redfield approach}
 \acro{BRT}{Bloch-Redfield tensor}
 \acro{BR}{Bloch-Redfield}
 \acro{SSA}{single shell approximation}
\end{acronym}

\section{Introduction}

\acp{QIS} were originally introduced to understand the local moment
formation in strongly correlated materials \cite{Anderson61} and the
screening of those moments via the Kondo effect
\cite{Kondo1964,KrishWilWilson80a,KrishWilWilson80b}.
The spectral properties of \acp{QIS} are of strong interest in several
different areas.  They are required in the framework of the \ac{DMFT}
\cite{Kuramoto85,Georges96,KotliarVollhardt2004} as part of the
self-consistent solution of the effective site problem.  They also
provide a fundamental understanding of the transport properties
through single electron transistors
\cite{KastnerSET1992,NatureGoldhaberGordon1998,goldhaberSET98,Kouwenhoven2000}
as well as playing a central role in interpreting \ac{STM} and \ac{STS}
of adatoms \cite{Manoharan2000,AgamSchiller2001} on surfaces. The
spectral properties carry information on the charge and spin dynamics
of molecules on surfaces
\cite{galperinNitzanRatner2006,MolecularElectronicsReview2009rohtua}
including inelastic processes
\cite{LorentePersson2000,REED2008,EickhoffSTM2020},
quantum phase transitions in the vicinity of graphene vacancies
\cite{Pareira2006,Cazalilla2012,MayGraphen2018,AndreiGraphen2018} and
in Ag-PTCTA dimer complexes \cite{AU-PTCDA-dimer}. These are only a
few examples in which the spectral information is crucial to understand
the properties of such systems.

In \acp{QIS} a small subsystem of interest with a small and finite
size Hilbert space is coupled to some free fermionic or bosonic baths
characterized by an energy continuum. The effect of these baths onto
the dynamics of the small subsystem is fully specified by the energy
dependent coupling functions. Many successful numerical applications
such as \ac{ED} \cite{CaffarelKrauth1994}, \ac{DMRG}
\cite{White92,Schollwoeck-2005,Schollwoeck2011} and the \ac{NRG}
approach \cite{Wilson75,BullaCostiPruschke2008} map the bath continuum
onto a discrete 1D tight-binding chain representation, and determine
an accurate solution of the finite size system. An exact solution of a
local spectral function in such a discretized finite size system is
always given by a set of $\delta$-peaks located at the elementary
excitation. To make contact with the correct continuum solution
of the original \ac{QIS}, a finite size broadening of the excitations
must be introduced in the Lehmann representation.  In the \ac{NRG}
this is typically done using a Gaussian broadening function
\cite{BullaCostiPruschke2008} on a logarithmic energy mesh with an
energy adapted artificial broadening parameter $b$.

In this paper, we present an approach for calculating spectral
functions using a hybrid \ac{NRG} method \cite{BoekerAnders2020}.  The
advantage of this method is the elimination of the artificial
broadening required by the standard approach.  It uses an exact
representation of the continuum bath coupling function to the impurity
by a combination of a finite size (Wilson) chain and augmenting
reservoirs coupled to each of the chain sites
\cite{SBMopenchain2017,BoekerAnders2020}.  Such a representation which
is based on a continuous fraction expansion was used
\cite{SBMopenchain2017} in the context of the \ac{SBM}
\cite{Leggett1987} to address the mass flow problem
\cite{VojtaBullaGuettgeAnders2010} close to critical coupling. While
in the application to the \ac{SBM} it was sufficient to include only
the real part of the reservoir correlation functions
\cite{SBMopenchain2017}, we recently presented a hybrid \ac{NRG}
approach to non-equilibrium dynamics \cite{BoekerAnders2020} that uses
the full complex reservoir correlation functions and thus includes
decoherence and dissipation. By combining the \ac{NRG} with a \ac{BRA}
we were able to induce a true thermalization and to show that the
steady-state is given by the equilibrium density matrix of the final
Hamiltonian, which is exact in the limits of the \ac{NRG}.

We modified the hybrid \ac{NRG} approach \cite{BoekerAnders2020} to
the calculation of spectral functions.  Starting from the exact
definition of the \ac{GF}, we investigate the time evolution of a
composite operator, consisting of the local creator and the density
matrix, in the presence of the reservoirs. By switching off all
reservoir couplings we recover the standard \ac{NRG} algorithm
\cite{PetersPruschkeAnders2006, WeichselbaumDelft2007} that is based
on using a complete basis set
\cite{AndersSchiller2005,AndersSchiller2006} for the \ac{NRG}
chain. The presence of the reservoir couplings leads to a master
equation for the interaction representation of the composite operator
when applying the secular approximation to the \ac{BRA}
\cite{MayKuehn2000,JeskeBRT2015}. As a consequence, the exact
excitations of the discretized system decay in different decay
channels which generates a finite life-time as well as a small Lamb
shift. In this way, we are generating a spectral function without the
need for an artificial broadening.

Although we derive an algorithm that is tailored to the \ac{NRG}
approach, the basic idea and the general scheme can also be adopted to
\ac{ED} or the \ac{DMRG} spectral function since here the continuum
representation of \acp{QIS} is typically obtained by a continued
fraction expansion as well \cite{RaasUhrigAnders04}. We expect that a
Krylov-space approach to spectral functions
\cite{NoceraKrylovDMRG2016} would be an ideal starting point to modify
our approach for \ac{ED} or the \ac{DMRG}.

\section{Green's functions in open quantum systems}
\label{sec:II}

\subsection{Introduction}

\acp{GF} play an important role for our understanding of the dynamics
and excitations of a system in and out of equilibrium. Here we focus
on the equilibrium case and restrict ourselves to single particle
\acp{GF} and linear susceptibilities.  There are three different types
of \acp{GF} for the operators $A,B$ that only dependent on the
relative time difference in equilibrium: The retarded \ac{GF}
\begin{eqnarray}
\label{eq:G-r-def}
G^\text{r}_{A,B}(t) &=& -i \Theta(t) {\rm Tr}\left[ \hat \rho [A(t), B]_s \right],
\end{eqnarray}
where $s=1$ for bosonic operators, $s=-1$ for fermionic operators $A,B$ and $
[A, B]_s= AB-sBA$, and on the other hand the lesser and greater
\ac{GF},
\begin{eqnarray}
G^{<}_{A,B}(t) &=& i {\rm Tr}\left[ \hat \rho A(t) B\right]
\\
G^{>}_{A,B}(t) &=& s i {\rm Tr}\left[ \hat \rho B A(t)\right] ,
\end{eqnarray}
where $\hat \rho$ is the density operator of the total system.  All
three \acp{GF} are related to the spectral function in
equilibrium. Therefore we focus on $G^\text{r}_{A,B}(t)$ and leave the
extension of the formalism to non-equilibrium to a future publication.

Two types of physical situations will be addressed in the general
theory below before we adapt the approach to quantum impurity models
solved with Wilson's \ac{NRG} approach
\cite{Wilson75,BullaCostiPruschke2008}. The first scenario is given by
a true open quantum system, for instance an electronic system where
the coupling to the quantized electromagnetic field induces a
radiative decay \cite{CarmichaelQuantumOpticsI}, while in the second
scenario involves a system with an excitation continuum that is approximatively
treated by some discretization scheme as used in \ac{ED}
\cite{CaffarelKrauth1994}, \ac{DMRG} \cite{White92,Schollwoeck-2005}
or \ac{NRG} \cite{Wilson75,BullaCostiPruschke2008} approaches.
In all these cases, the total Hamiltonian $H$ of the coupled problem is
divided into the part representing the system of interest, $H_{S}$,
the decoupled bath or reservoir $H_\text{R}$, and the coupling between
the two subsystems
$V=H_\text{SR}$:
\begin{eqnarray}
H &=& H_0 + V = H_{S}+H_\text{R} + H_\text{SR}.
\end{eqnarray}
While the Hamiltonians $H_{S}$ and $H_\text{R}$ are arbitrary, we
restrict the coupling between the two subsystems to be linear in the
elementary fermionic or bosonic operators, and $H_\text{SR}$ is taken
in the form
\begin{eqnarray}
V=H_\text{SR} &=& \sum_\alpha \lambda_\alpha O^{\rm S}_\alpha O^{\rm R}_\alpha
\end{eqnarray}
where $O^{\rm S(R)}_\alpha$ operates only on the system's (reservoir's) subspace.

In this paper, we consider \acp{QIS}.  Typically $H$ comprises a very
small system $H_{S}$ coupled to non-interacting bosonic of fermionic
reservoirs with continuous spectra. Successful numerical approaches
such as the \ac{DMRG} \cite{Schollwoeck-2005} or the \ac{NRG} map this
problem of infinitely many \ac{DOF} onto a 1D finite size chain
representation. Recently it was shown \cite{SBMopenchain2017,
BoekerAnders2020} that the full Hamiltonian $H$ can be recovered from
a Wilson chain representation of the problem by augmenting each chain
site with an auxiliary reservoir which can be analytically constructed
from the original problem.

\subsection{Derivation of a Bloch-Redfield approach adapted for Green's functions}
\label{sec:Bloch-Redfield-GFs}

The standard approaches
\cite{Dzhioev_2012,DoraArrigioni2015,LindbladContinuumVonDelft2016} to
the equilibrium and non-equilibrium \acp{GF} for open quantum systems
usually start from a Lindblad \cite{CarmichaelQuantumOpticsI}
extension to $H_{S}$ by parametrizing the couplings to the reservoir
by rates and Kraus operators. That has the advantage of ensuring the
positive definiteness of the reduced density operator at all times.
In this paper, however, we are interested in calculating the effect of
the full continuum spectra, which are neglected in a discretization
process, onto the dynamics of the system of interest. Therefore, we
use the \ac{BRA} \cite{MayKuehn2000} to explicitly derive the
relaxation constants from the original Hamiltonian instead of using
the Lindblad rates as fitting parameters \cite{DoraArrigioni2015}. By
making a fitting process superfluous, significantly more rate
parameters can be considered to model the effect of the
reservoirs. This constitutes the major difference of our method
compared to the recently proposed Lindblad approaches
\cite{DoraArrigioni2015,LindbladContinuumVonDelft2016} to restore the
continuum limit.

As observed in Ref.\ \cite{LindbladContinuumVonDelft2016}, all three
\acp{GF} introduced above require expectation values of the type
$\expect{ \hat \rho f(A(t),B)}$ where $f(x,y)$ usually is a linear
function in $x$ and $y$. The time evolution of $A(t)$ is determined by
the full Hamiltonian which commutes with $\rho$ in equilibrium. We
transform the expectation value such that a combination of $\rho$ and
$B$ becomes time dependent.  In case of $G^\text{r}_{A,B}(t)$,
\begin{eqnarray}
\label{eq:6}
{\rm Tr}\left[  \hat \rho [A(t), B]_s \right] &=& 
 {\rm Tr}\left[ A e^{-iHt} (B  \hat \rho- s \hat \rho B) e^{iHt} ) \right],
\end{eqnarray}
the  composite operator $ \hat\rho_{B}(t)$,
\begin{eqnarray}
\label{eq:rhoB-def}
 \hat\rho_{B}(t)&=& e^{-iHt} (B  \hat\rho- s \hat\rho B) e^{iHt},
\end{eqnarray}
 emerges
whose time evolution follows the von Neumann equation
\begin{eqnarray}
\partial_t O(t) &=& i [ O(t), H] .
\end{eqnarray}
Note that in case of the lesser or greater \ac{GF} we replace $O$ by
only the first or second term in Eq.\ \eqref{eq:rhoB-def}.

We can separate the operator $A$ that only acts on the Fock space of
$H_{S}$ from the operator $\rho_{B}(t)$ that acts on the total Fock
space of the coupled system.  In order to calculate the \ac{GF}, we
evaluate the trace in Eq.\ \eqref{eq:G-r-def} in two step:  first we
trace out the reservoir \ac{DOF} from the composite operator $
\hat\rho_{B}(t)$,
\begin{eqnarray}
 \hat\rho_{B,{S}} (t) &=& {\rm Tr}_\text{R}\left[ \hat \rho_{B}(t) \right]
\end{eqnarray}
to generate a reduced operator that only acts on the restricted Fock
space of $H_{S}$ and then we perform ${\rm Tr}_{S}\left[ A
\hat\rho_{B,{S}}(t) \right]$ in the second step.

We transform all operators into the interaction representation, i.\
e., $O(t)\to O^{I}(t)= \exp(iH_0 t) O(t) \exp(-iH_0t)$.  Following 
Appendix B of Ref.\ \cite{BoekerAnders2020}, we arrive at the
differential equation
\begin{align}
\label{eq:diff-equation-rho_B_S}
\partial_t  \hat \rho^{I}_{B,{S}} (t) &= - \int_0^t dt'
{\rm Tr}_\text{R}
\left[
[[   \hat\rho^{I}_\text{B}(t'), V^{I}(t')], V^{I}(t)]
\right],
\end{align}
where we made use of ${\rm Tr}_\text{R}[ [ \hat\rho^{I}_\text{B}(t),
V^{I}(t)] ]=0$ for particle number conserving interactions $V$.

To proceed, we employ the assumptions of the \ac{BRA}
\cite{MayKuehn2000}: (i) The interaction to the reservoirs, $V$, is
weak so that a second-order treatment is sufficient to obtain the
major contribution to the dynamics. (ii) There is no feedback of the
finite size system $H_{S}$ onto the infinitely large reservoirs such
that a decoupling $ \hat\rho^{I}_\text{B}(t) \approx
\hat\rho^{I}_{B,{S}} (t) \hat\rho_{R}$\cite{MayKuehn2000} is
justified. (iii) After eliminating the fast system dynamics by the
transformation into the interaction representation, the time dependent
operator $ \hat\rho^{I}_{B,{S}} (t)$ is only a slowly varying function
of time in comparison to the fast decay of the reservoir correlation
function on the relative time scale $\tau = t-t'$. Thus, we perform a
transformation $t'\to \tau=t-t'$ and replace $ \hat\rho_{B}(t-\tau)
\to \hat\rho_{B}(t)$ under the integral in Eq.\
\eqref{eq:diff-equation-rho_B_S}. Within this assumption, we can
replace the upper limit of the integral by $t \to \infty$.

The double commutator in Eq.\ \eqref{eq:diff-equation-rho_B_S} yields
four terms with different operator ordering of $
\hat\rho^{I}_\text{B}(t'), V^{I}(t')$ and $V^{I}(t)$.  For the
evaluation of these four terms we specify the form of the interaction
$V$. It was shown \cite{SBMopenchain2017,BoekerAnders2020} that the
continuum problem can be exactly restored by reservoir couplings of
the type
\begin{eqnarray}
\label{eq:def-H-SR}
H_\text{SR} &=& \sum_{\tilde m=0, \nu}^{N} t'_{\tilde m \nu} f^\dagger_{\tilde m \nu} c_{\tilde m\nu}^{ } + H.C.
\end{eqnarray}
where $f^\dagger_{\tilde m \nu}$ is a fermionic (bosonic) operator on
the Wilson chain site $\tilde m$, $\nu$ labels the flavor of a Wilson
chain \cite{Wilson75,BullaCostiPruschke2008} (which could resemble the
spin \ac{DOF}) 
and $c_{\tilde m \nu}$ denotes the fermionic (bosonic)
reservoir annihilation operator corresponding to the reservoir $\tilde
m \nu$. The value of $ t'_{\tilde m \nu}$ can be calculated with the
continuous fraction expansion outlined in the Refs.\
\cite{SBMopenchain2017,BoekerAnders2020}. For the further derivation,
only the analytic structure of $H_\text{SR}$ enters the equation so
that it is easy to adapt the rather general form of $H_\text{SR}$ to
other types of problems.

In the following, we focus on fermionic Wilson chains. By inserting
the form of $H_\text{SR}$ into Eq.\ \eqref{eq:diff-equation-rho_B_S}
we obtain almost the same differential equations for
$\rho^{I}_{B,{S}}(t)$ as for the reduced density matrix
\cite{BoekerAnders2020},
\begin{widetext}
\begin{eqnarray}
\label{eq:dgl-rho-red-B-S}
\partial_t  \hat\rho^{I}_{B,{S}}(t)  &=& 
 -
 i\sum_{\tilde m=0}^{N} \sum_\nu\int_0^\infty d\tau
  \hat\rho^{I}_{B,{S}}(t) \left[ f_{\tilde m \nu}^{ }(t-\tau) f^\dagger_{\tilde m \nu} (t) G^{>}_{\tilde m,\nu}(-\tau) - f_{ \nu \tilde m}^\dagger(t-\tau ) f_{\tilde m \nu}^{ } (t) G^{<}_{\tilde m,\nu}(-\tau)  \right]
\non
&& + is\sum_{\tilde m=0}^{N} \sum_\nu
\int_0^\infty d\tau 
\left[ f^\dagger_{\tilde m \nu}(t-\tau)  \hat \rho^{I}_{B,{S}}(t)  f_{\tilde m \nu}(t) G^{>}_{\tilde m,\nu}(\tau) - f_{\tilde m \nu}(t-\tau )  \hat \rho^{I}_{B,{S}}(t)  f_{\tilde m \nu}^\dagger (t) G^{<}_{\tilde m,\nu}(\tau)  \right]
\non
&& + is\sum_{\tilde m=0}^{N} \sum_\nu
\int_0^\infty d\tau
\left[ f^\dagger_{\tilde m \nu}(t)   \hat\rho^{I}_{B,{S}}(t)  (t)  f_{\tilde m \nu}(t-\tau) G^{>}_{\tilde m,\nu}(-\tau) - f_{\tilde m \nu}(t )  \hat \rho^{I}_{B,{S}}(t)  f_{\tilde m \nu}^\dagger (t-\tau) G^{<}_{\tilde m,\nu}(-\tau)  \right]
\non
&&
 - i\sum_{\tilde m=0}^{N} \sum_\nu\int_0^\infty d\tau
 \left[ f^{ }_{\tilde m \nu}(t) f_{\tilde m \nu}^\dagger(t-\tau) G^{>}_{\tilde m,\nu}(\tau) - f_{\tilde m \nu}^\dagger(t) f_{\tilde m \nu}^{ } (t-\tau) G^{<}_{\tilde m,\nu}(\tau)  \right]  \hat\rho^{I}_{B,{S}}(t),
\end{eqnarray}
\end{widetext}
where the complex reservoir  correlation functions are defined as
\begin{eqnarray}
\label{eq:G-greater}
G^{>}_{\tilde m,\nu}(\tau) &=& -i|t'_{\tilde m \nu}|^2
{\rm Tr}_\text{R} 
\left[ \hat\rho_{R}
c^\dagger_{\tilde m \nu}(\tau) c_{\tilde m \nu}
\right]
\\
\label{eq:G-lesser}
G^{<}_{\tilde m,\nu}(\tau) &=& i|t'_{\tilde m \nu}|^2
{\rm Tr}_\text{R}
\left[ \hat\rho_{R}
c_{\tilde m \nu}(\tau)   c^\dagger_{\tilde m \nu}
\right].
\end{eqnarray}
We introduced the lesser and greater \acp{GF} at this stage to make a
connection to other approaches in the literature.  The more convenient
way of writing the equations would be by omitting the complex factor
$i$ in  Eqs.\ \eqref{eq:dgl-rho-red-B-S} -\eqref{eq:G-lesser}.

The difference between the standard \ac{BR} equations for the density
matrix and our approach arises from the terms that give rise to the second and third
line in Eq.\ \eqref{eq:dgl-rho-red-B-S}, i.e.  terms of the form
$V^{I} \hat \rho V^{I}$. An additional sign $s=-1$ is generated from
the commutation of $ \hat\rho^{I}_{B,{S}}(t)$ with the reservoir
fermionic operators for a fermionic \ac{GF}.  The operator dynamics is
evaluated by converting the equation into a matrix equation in the
eigenbasis of $H_{S}$: Then the time evolution of the operators
$f_{\tilde m \nu }(t)$ becomes trivial, and the remaining time integral can be
calculated easily.  Details can be found in Ref.~\cite{MayKuehn2000}.  
After solving this expression, we end up with a
\ac{BR} type master equation that contains the information about the
decay of the system eigenstates due to the coupling to the reservoirs.

This first part of the approach can be adapted to any system of
interest. After eliminating the time-dependency the structure is
related to the Lindblad approach used in the literature
\cite{CarmichaelQuantumOpticsI,Dzhioev_2012,DoraArrigioni2015,LindbladContinuumVonDelft2016}.

\subsection{Quantum impurity systems and the Numerical Renormalization Group}

\subsubsection{Wilson chain representation and augmented reservoirs}

To apply the approach outlined in 
Sec.~\ref{sec:Bloch-Redfield-GFs} to the \acp{GF} of \acp{QIS} we
introduce the notations and summarize the basic ideas of a complete
basis set \cite{AndersSchiller2005,AndersSchiller2006} originally
developed for the non-equilibrium extension of the \ac{NRG}.

The Hamiltonian of a \ac{QIS} comprise the local dynamics $H_{\rm
imp}$, a set of non-interacting baths $H_{\rm bath}$ and a coupling
between these two subsystems $H_{I}$:
\begin{eqnarray}
\label{eq:QIS}
H &=H_{\rm imp} +H_{\rm bath} + H_{I} .
\end{eqnarray}
$H_{\rm bath}$ accounts for non-interacting and continuous baths
\begin{eqnarray}
H_{\rm bath} &=& \sum_{\nu} \sum_{k} \e_{k\nu}^{ } c^\dagger_{k\nu} c_{k\nu}^{ },
\end{eqnarray}
counted by a flavor index $\nu$ (combined spin and band index). The
operator $c^\dagger_{k\nu}$ creates a free bath electron of flavor
$\nu$ and wave number $k$ with the energy $\e_{k \nu}$.  The coupling
$H_{I}$ between the two subsystems is parameterized by a general
hybridization term
\begin{eqnarray}
\label{eqn:Hhyp}
H_{I} &=& \sum_{\nu}V_\nu  \left(c^\dagger_{0\nu} A_\nu^{ } + A^\dagger_{\nu} c_{0\nu}^{ }\right)
\end{eqnarray}
where we follow the notation of Ref.\ \cite{BoekerAnders2020}.  The
operator $c_{0\nu}$ annihilates a local bath state of flavor $\nu$
defined as a linear combination of annihilators $ c_{k\nu}$ of bath
modes with the eigenenergy $\e_{k\nu}$
 \begin{eqnarray}
\label{eqn:c-local-orig}
c_{0\nu} &=& \sum_{k} \lambda_{k\nu} c_{k\nu}
\end{eqnarray}
such that $c_{0\nu}$ fulfills canonical commutation relations.
$A^\dagger_\nu (A_\nu)$ creates (annihilates) a local fermionic
excitation in the Hilbert space of $H_{\rm imp}$ and is given by a
linear combination of local impurity orbitals. The coupling parameters
$\lambda_{k\nu}$ contain the possible energy-dependent hybridization.
The baths influence on the local dynamics of the impurity \ac{DOF}
is fully determined
\cite{Wilson75,Leggett1987,BullaPruschkeHewson1997} by the coupling
function $\Delta_\nu(z)$ defined as
\begin{eqnarray}
\label{eqn:local-hyb}
\Delta_\nu(z) &=& V^2_{\nu} \sum_k \frac{\lambda^2_{k\nu}}{z-\e_{k\nu}}
\punkt
\end{eqnarray}

Using the control parameter $\Lambda > 1$, the Hamiltonian Eq.\
\eqref{eq:QIS} is approximated by a Wilson chain representation in the
\ac{NRG} \cite{Wilson75,BullaCostiPruschke2008},
\begin{eqnarray}
H_N^{\rm NRG} &=& H_{\rm imp}  + H_{I} + H_{\rm chain}(N),
\label{eqn:h-nrg-n}
\end{eqnarray}
where
\begin{eqnarray}
\label{eqn:H-chain-N}
H_{\rm chain}(N) &=&
\sum_{m=0}^{N} \sum_\nu \e_{m\nu} f^\dagger_{m\nu} f_{m\nu} 
\\
&&
+ \sum_{m=1}^{N}\sum_\nu t_{ m-1  \nu}\left(f^\dagger_{m\nu} f_{m-1\nu} 
+f^\dagger_{m-1\nu} f_{m\nu}\right),
\nonumber
\end{eqnarray}
and an unaltered $H_{\rm imp}$.  The sequence of Hamiltonians
$H^\text{NRG}_m$ is iteratively diagonalized, discarding the
high-energy states at each step to maintain a manageable number of
states.  The original Hamiltonian could be recovered
\cite{Wilson75,BullaCostiPruschke2008} by
\begin{eqnarray}
H &=& \lim_{\Lambda\to 1^+}\lim_{N\to\infty}H_{N}^{\rm NRG}(\Lambda).
\end{eqnarray}
Note that the hopping matrix elements in Eq.\ \eqref{eqn:H-chain-N},
$t_{ m  \nu}$, scale as $t_{ m  \nu}\propto \Lambda^{-m/2}$. Therefore $H_{N}^{\rm NRG}(\Lambda)$
in Eq.\ \eqref{eqn:H-chain-N} is given in absolute units.

For any finite size representation $H_N^{\rm NRG}$, 
the full continuum limit  of the model can be
recovered if the approximated bath representation $H_{\rm bath} \to H_{\rm chain}(N)$
is augmented by a set of 
additional reservoirs 
\begin{eqnarray}
H_{\rm res}(m) &=& \sum_{\nu} \sum_{k} \e_{m k\nu}^{ } c^\dagger_{m k\nu} c_{m k\nu}^{ },
\end{eqnarray}
each coupled to the respective $m$-th site of the tight-binding chain via
\begin{eqnarray}
\label{eq:H_I_N}
H_{I}(m) &=& \sum_{\nu} t'_{m \nu}  \left(c^\dagger_{m \nu} f_{m \nu}^{ } + f^\dagger_{m \nu} c_{m \nu}^{ }\right)
\end{eqnarray}
such that
\begin{eqnarray}
\label{eqn:hybrid-nrg-hamiltionian-N-reservoirs}
H_{\rm bath} 
&=& H_{\rm chain}(N) + \sum_{m=0}^N \left( H_{\rm res}(m) + H_{I}(m) \right)
\end{eqnarray}
yields the same local impurity dynamics as the original model.  
Analogously  to $H_\text{bath}$, $H_{\rm res}(m)$ is a reservoir of fermionic (bosonic) modes
coupling to the m-th chain site of the Wilson chain. 
For $m<N$, these modes are restricted to the high-energy modes which are neglected in the standard Wilson chain. For further
details on the construction of the reservoir coupling functions see
references \cite{SBMopenchain2017,BoekerAnders2020}. Note that the
coupling between the chains and the reservoirs, $\sum_{m=0}^N H_I(m)$,
takes on the role of $H_\text{SR}$ in the previous section.

\subsubsection{Complete basis set}

It has been proven \cite{AndersSchiller2005,AndersSchiller2006} that
the set of eigenstates of $H^\text{NRG}_{m}$, $\{\ket{l,e;m}\}$,
discarded after each \ac{NRG} iteration 
partitions the many-body
Fock space and defines a complete basis set
\begin{eqnarray}
\sum_{m = m_{\rm min}}^N \hat P_\text{d}^{(m)} &=& \hat 1 .
\label{equ:complete-basis}
\end{eqnarray}
where $m_\text{min}$  is defined as the first NRG iteration at which states are discarded.
In the notation $\{\ket{l,e;m}\}$, $l$ labels all discarded states after iteration $m$
and the variable $e$ denotes the number operator basis of the
remaining degrees of freedom of the Wilson chain from chain site $m+1$ until $N$.
For more details see Refs.\ \cite{AndersSchiller2005,AndersSchiller2006}.
This basis set is also an approximate eigenbasis
of the full Hamiltonian.
Here we have defined 
\begin{eqnarray}
\hat P_\text{d}^{(m)} &=& \sum_{l,e}      \ket{l,e;m}\bra{l,e;m} 
\label{equ:complete-basis_Pd}
\end{eqnarray}
as the projector onto the subspace spanned by the discarded states after the iteration $m$ and likewise
\begin{eqnarray}
\hat P_\text{k}^{(m)} &=& \sum_{k,e}      \ket{k,e;m}\bra{k,e;m} 
\label{equ:complete-basis_Pk}
\end{eqnarray}
as the projector onto all retained states after the iteration $m$. For
discarded states we use the index $l$, while for kept states the index
$k$ is used. The environment variable $e$ accounts for the tensor
product basis of the remaining chain sites $m+1,m+2,\cdots N$.

The complete basis set entering Eq. \eqref{equ:complete-basis} has
been used to calculate non-equilibrium dynamics
\cite{AndersSchiller2005,AndersSchiller2006} in \acp{QIS} as well as
deriving a sum-rule conserving representation of spectral functions
\cite{PetersPruschkeAnders2006,WeichselbaumDelft2007}.
Alternatively, we can also focus on a specific iteration $m$. Then the
Fock space can be partitioned by two complementary projection
operators $\hat 1_m^-$ and $\hat 1_m^+$:
\begin{eqnarray}
\hat 1_m^- &=& \sum_{m'=m_{min}}^{m} \hat P_\text{d}^{(m')} \; ,
\label{eqn:partition-fock-space-i}
\\
\hat 1_m^+ &=& \sum_{m'= m+1}^N \hat P_\text{d}^{(m')} = \hat P_\text{k}^{(m)}
\label{eqn:partition-fock-space-ii}
\end{eqnarray}
with the  completeness relation
\begin{eqnarray}
  \hat 1 &=& \hat{1}_m^- + \hat{1}_m^+.
\label{eqn:completness}
\end{eqnarray}
Note that for $m = N$ only $\hat 1_m^-$ exists since all states are considered discarded after the last iteration.

Let us partition a generic operator $\hat O$ in the sectors of the complete Fock space
by employing the  completeness relations \eqref{equ:complete-basis} and \eqref{eqn:completness}.
\begin{eqnarray}
\hat O &=& \hat 1 \hat O \hat 1
= 
 \sum_{m=m_{min}}^N \hat P_\text{d}^{(m)}  \hat O \left( \hat 1^-_m +  \hat P_\text{k}^{(m)} \right)
\non
&=&
 \sum_{m=m_{min}}^N \hat P_\text{d}^{(m)}  \hat O   \hat P_\text{k}^{(m)}
  +
 \sum_{m=m_{min}}^N \hat P_\text{d}^{(m)}  \hat O  \hat P_\text{d}^{(m)}
 \non &&
+
 \sum_{m=m_{min}}^N \hat P_\text{d}^{(m)}   \hat O  \left( \sum_{m'=m_{min}}^{m-1} \hat P_\text{d}^{(m')} \right).
 \end{eqnarray}
By rearranging the summation of $m,m'$ in the last term and using
Eq.~(\ref{eqn:partition-fock-space-ii}), we obtain

\begin{eqnarray}
\sum_{m=m_{min}}^N \hat P_\text{d}^{(m)}   \hat O  \left( \sum_{m'=m_{min}}^{m-1} \hat P_\text{d}^{(m')} \right)
\non
= \sum_{m=m_{min}}^N \hat P_\text{k}^{(m)}   \hat O  P_\text{d}^{(m)}.
\label{eqn:red_m}
\end{eqnarray}
Therefore, the operator $\hat O$ is given
by the exact representation
\begin{eqnarray}
\hat O &=&
 \sum_{m=m_{min}}^N  \hat O_\text{d}(m)
 \label{eqn:24}
  \end{eqnarray}
where the  part $\hat O_\text{d}(m)$ consists of the three terms:
\begin{eqnarray}
 \label{eqn:def-O-d}
\hat O_\text{d}(m) &=& 
 P_\text{d}^{(m)}  \hat O  \hat P_\text{d}^{(m)}
 +
  P_\text{d}^{(m)}  \hat O  \hat P_\text{k}^{(m)}
+
 P_\text{k}^{(m)} \hat O  \hat P_\text{d}^{(m)}.
\end{eqnarray}
The first term remains diagonal in the subspace spanned by the
discarded states at iteration $m$ while the two others describe
excitations between the sector of discarded and the sector of kept
states. We recognize the structure of the operators already known from
the time-dependent \ac{NRG}
\cite{AndersSchiller2005,AndersSchiller2006}: At each iteration, or
energy scale, $m$ denotes only the discard-discard or kept-discarded parts of
the operator matrix contribution while the respective kept-kept part is
refined at a later iteration.

\subsection{NRG approach to  Green's functions}

The equilibrium real-time retarded \ac{GF} is defined in Eq.\
\eqref{eq:G-r-def}. In the \ac{NRG}, we are typically only interested
in \acp{GF} for local operators $A,B$ that are diagonal in the
environment variable $e$ of the complete basis.

To review the established \ac{NRG} approach to spectral functions
\cite{PetersPruschkeAnders2006,WeichselbaumDelft2007}, we neglect
$ H_I$ and restrict ourselves to solutions of the Wilson chain
representations.  Since the expansion of a local operator as
introduced in Eq.\ \eqref{eqn:24} becomes diagonal in the environment
variable $e$, Eq.\ \eqref{eq:6} can be written as
\begin{align}
\label{eq:nrg-gf-def-36}
{\rm Tr}\left[  \hat \rho [A(t), B]_s \right] =
\sum_{m=m_{min}}^N \sum_{r,s}^{\rm trunc}  A_{r,s}(m) e^{i(E^m_r-E^m_s)t}
\nonumber
\\
\times \sum_e \bra{s,e;m}  \hat\rho^{I}_{B,{S}}(0) \ket{r,e;m},
\end{align}
where $r,s$ must contain at least one discarded state according to
Eq.~\eqref{eqn:def-O-d}. $E^m_r$ denotes the NRG eigenenergy of the
eigenstate $\ket{r,e;m}$ at the NRG iteration $m$. The trace over the
environment only acts on the operator $
\hat\rho^{I}_{B,{S}}(0)$. Since we restrict to the decoupled problem,
$V^{I}(t)=H^{I}_\text{C-R}(t)=0$, $\rho$ on the left hand side
factorizes as $ \hat\rho= \hat\rho_{S} \hat\rho_{R}$ and only the
thermodynamics density operator of the Wilson chain, $\rho_{S}$, is
entering the calculation of the trace in Eq.\
\eqref{eq:nrg-gf-def-36}.  Consequently, we can replace
$\rho^{I}_{B,{S}}(t)$ by its initial value. The exponential phase
factor, $e^{i(E_r-E_s)t}$, emerges from the transformation of
$\rho^{I}_{B,{S}}(0)$ back into the original Schr\"odinger
representation.

We explicitly used the fact that the basis states are approximate
eigenstates of the \ac{NRG} Hamiltonian $H_N^{\rm NRG}$. We only need
to calculate the reduced density matrix $\rho^{(B)}_{s,r}(m)$. Since
the operator $B$ is also diagonal in $e$, we arrive at
\begin{align}
{\rm Tr}\left[ \rho [A(t), B]_s \right] =
\sum_{m=m_{min}}^N \sum_{r,s}^{\rm trunc} \sum_{s'}  A_{r,s}(m) e^{i(E^m_r-E^m_s)t}
%e^{i(E_r-E_s)t} 
\nonumber
\\
\times \left[ B_{s,s'}(m) \rho^{\rm red}_{s',r}(m) -s \rho_{s,s'}^{\rm red} (m)B_{s',r}(m)
\right] 
\label{eq:nrg-gf}
\end{align}
where the reduced density matrix is defined as
\begin{eqnarray}
\rho_{r',r}^{\rm red}(m) &=& \sum_{e} \bra{r',e;m} \hat \rho_{S}(0) \ket{r,e;m},
\label{eq:red_dens_mat_T0}
\end{eqnarray}
where at least one element of the index pair $(r,s)$ also must be a
discarded state at iteration $m$.  If the \ac{NRG} density matrix
\begin{eqnarray}
\label{eqn:rho-nrg}
\hat \rho_{S}(0) &=& \frac{1}{Z_N} \sum_{l} e^{-\beta E_l^N}  \ket{l;N} \bra{l;N},
\end{eqnarray}
with $Z_N= \sum_l \exp(-\beta E_l^N)$ is used $s'$ in Eq.\
\eqref{eq:nrg-gf} must be a kept state for $m<N$. On the other hand,
if the full density matrix \cite{WeichselbaumDelft2007} is used $s'$
runs over both kept and discarded states of iteration $m$.

Fourier transformation of this expression leads to the Lehmann
representation of the spectral function that is found in the Refs.\
\cite{PetersPruschkeAnders2006,WeichselbaumDelft2007}.  Consequently,
we can interpret the \ac{NRG} representation, Eq.~\eqref{eq:nrg-gf},
as the finite size approximation of the original continuum problem by
setting the coupling $ H_I(N)=0$.

\subsection{Bloch-Redfield approach to NRG Green's functions}
\label{subsec:BRA}

To recover an approximate solution of the continuum problem
we perturbatively include $ H_I(N)$. The starting point is the
interaction representation of $\rho_{B}$.  Substituting $\rho_{B}(t) =
\exp(-iH_0 t) \rho^{I}_\text{B}(t) \exp(iH_0 t)$ into Eq.\
\eqref{eq:6} yields the same expression as Eq.\
\eqref{eq:nrg-gf-def-36} but with the replacement $\rho_{B,{S}}(0)\to
\rho^{I}_{B,{S}}(t)$: We are left with calculating the dynamics of the
corresponding reduced composite operator
\begin{eqnarray}
\rho^{B,\rm red}_{r,s}(m;t) &=& \sum_{e}  \bra{r,e;m} \rho^{I}_{B,{S}} (t)\ket{s,e;m}.
\end{eqnarray}

As noted before, Eq.\ \eqref{eq:dgl-rho-red-B-S} has the same analytic
structure as the dynamics of the reduced density matrix $\rho_{S}(t)$
without the additional operator $B$, with the exception of the sign change
$s=-1$ for fermionic GFs. After transforming Eq.\
\eqref{eq:dgl-rho-red-B-S} into a matrix representation using the
complete basis set of the \ac{NRG}, we can make use of the results of
Ref.\ \cite{BoekerAnders2020}.  Under the time integral, rapidly
oscillating terms of the type $e^{i E t}$ occur, which are assumed to
only have a contribution if $E=0$. This is a secular approximation,
and effectively replaces the oscillating terms by Kronecker-deltas.
One ends up with the rate equation
\begin{align}
\dot{\rho}^{B,\rm red}_{r_1,r_2}(m;t) = - \sum_{r_3,r_4} 
R_{r_1,r_2; r_3,r_4} (m)
\rho_{r_3,r_4}^{B,\rm red}(m;t),
\label{ODDMMasterEqu}
\end{align}
where $R_{r_1,r_2; r_3,r_4} (m)$ denotes the time-independent \ac{BRT}
which can be calculated from Eq.\ \eqref{eq:dgl-rho-red-B-S} and is
given by the expression \cite{BoekerAnders2020}
\begin{widetext}
\begin{subequations}
\label{MasterEqu}
\begin{align}
% R^B_{r_1,r_2;r_3,r_4} (m) 
R_{r_1,r_2;r_3,r_4} (m)
&= \delta_{r_2,r_4} \sum_{r_5} \Xi_{r_1,r_5;r_5,r_3} (m) + \delta_{r_1,r_3} \sum_{r_5} \Xi^*_{r_4,r_5;r_5,r_2} (m) - 2 s \Re \Xi_{r_4,r_2;r_1,r_3} (m)
\label{MasterEqu-b}
\\
\Xi_{r_1,r_2;r_3,r_4} (m) &= 
\delta_{\omega_{1,2}+\omega_{3,4},0}
\sum_{\tilde{m}=0}^{m} \sum_\nu \big[C_{\nu,\tilde{m}}(\omega_{3,4})(f^\dag_{\tilde{m}\nu})_{1,2} (f^{ }_{\tilde{m}\nu})_{3,4} +\bar C_{\nu,\tilde{m}}(\omega_{3,4})(f^{ }_{\tilde{m}\nu})_{1,2} (f^{\dag}_{\tilde{m}\nu})_{3,4} \big] ,
\label{MasterEqu-c}
\end{align} 
\end{subequations}
\end{widetext}
where $(f^{(\dag) }_{\tilde{m}\nu})_{i,j}=\bra{r_i;m}
f^{(\dag)}_{\tilde{m}\nu}\ket{r_j;m} $ denotes the shortcut
notation of the $\tilde m$-th chain site operators.  Note that the
latter matrix elements are diagonal in the environment $e$ such that
the traces of the environment variables have been performed leading to
the reduced matrix elements $\rho_{r,r'}^{B,\rm red}(m;t)$.  The
energy differences $\omega_{i,j}$ are defined as $\omega_{i,j}=
E_{r_i}^m-E^m_{r_j}$, and the coupling strength to the additional
reservoirs is encoded into the half-sided Fourier transformation
\cite{BoekerAnders2020},
\begin{subequations}
\label{eq:reservoir-gfs-FT}
\begin{eqnarray}
C_{\tilde m,\nu}(\w) &=& i\int_0^{\infty} d\tau G^{>}_{\tilde m,\nu}(\tau) e^{-i\w \tau}\\
\bar C_{\tilde m,\nu}(\w) &=& -i\int_0^{\infty} d\tau G^{<}_{\tilde m,\nu}(\tau) e^{-i\w \tau}  .
\end{eqnarray}
\end{subequations}
Note that we distinguish two different indices $m$ and $\tilde
m$. While $m$ denotes the \ac{NRG} iteration, $\tilde m$ labels the
reservoir coupled to the Wilson chain site $\tilde m$.

The major difference to the standard \ac{BRT}
\cite{MayKuehn2000,JeskeBRT2015} are the additional signs $s$ in the
last two terms of the relaxation rates in Eq.\ \eqref{MasterEqu-b}
that depend on the statistics of the operator $B$.

A comment is warrented 
regarding the difference between the \ac{BRA} and
the evaluation of a Lindblad equation
\cite{CarmichaelQuantumOpticsI,DoraArrigioni2015}.  The general form
of the \ac{BR} equation, Eq.\ \eqref{eq:dgl-rho-red-B-S}, does not
guarantee that the density operator maintains its positive
definiteness: The approach requires a consistent microscopic noise
model \cite{JeskeBRT2015}.  In the secular approximation, the \ac{BR}
equation can be mapped onto a Lindblad form when we additionally
replace $C_{\nu,\tilde{m}}(\omega)$ by a constant. Furthermore, there
is a debate in the literature \cite{JeskeBRT2015} whether the full
secular approximation is justified. In our case, however, only
off-diagonal matrix elements of $\rho^{B,\rm red}_{r_1,r_2}(m;t)$ are
non-zero, and therefore this debate is irrelevant here.

\subsubsection{Calculation of the contributions to the \ac{BRT}}
\label{SS:contr_to_BRT}

I addition to the approximations mentioned in
Sec. \ref{sec:Bloch-Redfield-GFs}, as well as the secular
approximation, we have applied further assumptions to arrive at Eq.\
\eqref{ODDMMasterEqu}. For a more elaborate overview see
Ref. \cite{BoekerPhD2021}.

As explained above, the complete basis set in principle covers the
Fock space of all \ac{NRG} iterations, for which high-energy states
are discarded [see Eq.\ \eqref{eqn:completness}]. Consequently, Eq.\
\eqref{ODDMMasterEqu} should directly include four independent sums
over the discarded states of all \ac{NRG} iterations plus indirectly
include a sum for $m_5$ and $\tilde m$ in Eq.\ \eqref{MasterEqu-b} and
Eq.\ \eqref{MasterEqu-c}, respectively.

By using the partitioning Eq.\ \eqref{eqn:completness}, the indices
can be reduced to $m_1=m_2=m$ and $m_3=m_4=m'$ in the manner of Eq.\
\eqref{eqn:red_m}. This is still an exact result and in that way the
\ac{BRT} in the rate equation \eqref{ODDMMasterEqu} couples the
reduced composite matrix $\rho_{r,s}^{B,\rm red}(m;t)$ of two
different \ac{NRG} iterations $m$ and $m'$ Consequently, the operators
$\Xi_{r,s;r',s'}(m,m')$ include two different iterations as well.

The secular approximation requires $E_{r_1}^m = E_{r_3}^{m}$ for the
first term on the r.h.s of Eq.\ \eqref{MasterEqu-b} as well as
$E_{r_4}^m = E_{r_2}^{m}$ for the second term. Using the NRG hierarchy
and excluding accidental energy degeneracies leads to the replacement
$\delta_{r_1,r_3} \delta_{r_2,r_4} \sum_{r_5} \left(
\Xi_{r_1,r_5;r_5,r_1} (m) + \Xi^*_{r_2,r_5;r_5,r_2} (m) \right)$ for
these two contributions.  This solely defines the diagonal part of the
\ac{BRT} where $r_1 = r_3, r_2 = r_4$ holds.

The sum running over the states $r_5$ in Eq.\ \eqref{MasterEqu-b},
however, still involves discarded states of NRG iterations $m_5 <
m$. Reminding ourselves that $T \approx \w_N$ in the NRG, where
$\w_N\propto \Lambda^{-N/2}$ denotes the characteristic energy scale
of the last NRG iteration, excitation energies $\w = E^{m_5}_{r_5} -
E^m_r\gg k_B T$ are always positive for $m_5<m$. Since $\Re
C_{\nu,\tilde{m}}(\omega) = \Gamma_{\nu,\tilde{m}}(\omega) f(\w)$, the
Fermi-function $f(\w)$ exponentially suppresses such
contributions. Consequently, restricting the $r_5$ summation to the
shell $m_5=m$ has only very small impact on the life-time. This,
however, does not hold for $\Im C_{\nu,\tilde{m}}(\omega)$. Thus, the
neglect of contributions from $m_5 < m$ imposes a significant effect
on the imaginary part of the \ac{BRT} and consequently on the
resulting Lamb-shift in the spectrum. The Lamb-shift itself is small
for a particle-hole symmetric reservoir, which justifies the
approximation in our case. To reconstruct the correct Lamb-shift for a
strongly particle-hole asymmetric bath spectrum, a full calculation of
the diagonal elements of the \ac{BRT} is required, which is more
tedious but practically possible.  In Ref. \cite{BoekerAnders2020} we
have already implemented the case $|m-m'| \leq 1, \tilde m \leq
\text{min}\{m,m' \}$. A detailed instruction can be found in
Ref. \cite{BoekerPhD2021}.

This brings us to the second approximation, i.e. the so called local
operator approximation. Here we neglect all reservoirs $\tilde m >
\text{min}\{m,m' \}$, which lets us interpret the chain operators as
being "local" with respect to the environment \ac{DOF} $e , e'$ of the
iterations $m, m'$. This approximation simplifies the calculation of
the chain operators by enabling one to obtain them "on the fly" during
the course of the \ac{NRG} procedure, and is mainly justified by the
fact, that the contribution of $\Gamma_{\nu,\tilde{m}}(E^{m'}_{r'} -
E^m_r)$ is exponentially small, if $\tilde m > \text{min}\{m,m'
\}$ \cite{BoekerAnders2020}. Note however, that this approximation is,
just like the first one, not crucial for the application of our
method.

The last term on the r.h.s of Eq.\ \eqref{MasterEqu-b} is zero in the
case of $r_1 = r_3, r_2 = r_4$, since $f^\dag_{\nu,\tilde{m}}$ has only off-diagonal matrix elements. 
Therefore this term does not contribute to the diagonal part of the
\ac{BRT} in the superoperator space, i.\ e.\ to $R_{r_1,r_2;r_1,r_2}$.
On the contrary, the remaining two terms only have a contribution to $R_{r_1,r_2;r_1,r_2}$.
Hence, the diagonal part of the \ac{BRT} equals the first two terms, while the off-diagonal 
part of the \ac{BRT} is solely defined by the last term on the r.h.s of Eq.\ \eqref{MasterEqu-b}. 
Furthermore, since this term is real it has no effect on the
Lamb-shift. Due to the Kronecker-delta, it is only non-zero for index
combinations that  fulfill $E^{m}_{r_1} - E^{m}_{r_2} = E^{m'}_{r_3} -
E^{m'}_{r_4}$. For the calculation of spectral functions we focus on
finite energy excitations $|E^{m}_{r_1} - E^{m}_{r_2}|>0$.  Since
fermionic and bosonic creation operators connect different sectors of
the Fock space, the number of pairs $(r_1,r_2)$ and $(r_3,r_4)$ that
have the same non-zero excitation energies are very limited. This is
the justification to demand $m = m'$, which is called the \ac{SSA}. It
leads to $N - m_\text{min}$ entirely separate \acp{BRT}, which are
diagonalized independently of each other on the backwards run of the
\ac{NRG} procedure, as was implemented in
Ref. \cite{BoekerAnders2020}. The \ac{SSA} is essential for the
practical implementation of our \ac{BRA} to spectral functions, since
the construction and diagonalization of the large \ac{BRT} coupling 
all \ac{NRG} iterations is unfeasible.

In contrast to the density matrix, the diagonal matrix elements of the
composite operator, $\rho_{r,r}^{B,\rm red}(m;t)$, are zero and thus
the \ac{SSA} here does not violate essential physical properties as,
e.g., the conservation of the trace (see
Ref. \cite{BoekerAnders2020}).

\subsubsection{Application to the Green's function}
\label{SS:Appl_to_GF}

For the case of a vanishing temperature $T \to 0$ and a sufficiently
long Wilson chain, we can apply another approximation. As mentioned
above, here the Fermi-function cuts off all positive energy
arguments. Since the off-diagonal part of the \ac{BRT} is proportional
to the Fermi-function, the \ac{BRT} effectively becomes a tridiagonal
matrix, which means, that the off-diagonal elements do not impact the
eigenvalues of this matrix. Furthermore, due to the distinct shape of
the composite operator $\rho_{r,s}^{B,\rm red}(m;t=0)$, these elements
effectively do not contribute to the rate equation\
\eqref{ODDMMasterEqu} at all.
In this case the solution of the density matrix dynamics is
analytically given by
\begin{eqnarray}
\rho^{B,\rm red}_{r_1,r_2}(m;t) &=& e^{-R_{r_1,r_2; r_1,r_2} (m)t} \rho^{B,\rm red}_{r_1,r_2}(m;0)
\label{eq:BRT_diag}
\end{eqnarray}
and the complex tensor element $R_{r_1,r_2; r_1,r_2} (m)$ contains the
decay rate as well as, via its imaginary part, the Lamb-shift of the
excitation energy.  Substituting this decay matrix element back into
Eq.\ \eqref{eq:nrg-gf} for the \ac{NRG} \ac{GF} yields a modification
of the time dependency
\begin{eqnarray}
e^{i(E^m_r-E^m_s)t} &\to & e^{i(E^m_r-E^m_s)t} e^{-R_{s,r;s,r}t}, 
\end{eqnarray}
implying a contribution
\begin{align}
\frac{1}{\w - (E^m_r-E^m_s + \Im R_{s,r;s,r})  + i \Re R_{s,r;s,r}},
\label{equ:T0_Lor}
\end{align}
to the retarded \ac{GF} in the frequency domain. Note that $\Re
R_{s,r;s,r}>0$, since the zero eigenvalue of the \ac{BRT} can only
occur in the $r_1 = r_2$ subspace, where $\rho^{B,\rm
red}_{r_1,r_1}(m;t) = 0$.  Each pole of the original Lehmann
representation acquires a (small) Lamb-shift $\Im R_{s,r;s,r}$ (which
we will neglect in this paper), as well as a line width $\Re
R_{s,r;s,r}$. A closer inspection of Eq.\ \eqref{MasterEqu} leads to
the form $R_{s,r;s,r} =\chi_s+\chi^*_r$ with $\chi_s = \sum_{s'}
\Xi_{s^{ }, s';s',s^{ }}(m)$, which can be used for a significant
speed up of the calculation for $T \to 0$.

In this paper, we assumed the equilibrium density matrix of the Wilson
chain to be given by Eq.\ \eqref{eqn:rho-nrg}. Our method,
however, can be extended to the full density matrix approach
\cite{WeichselbaumDelft2007}, where the off-diagonal elements of the
\ac{BRT} have to be considered as well. We define a subset of
eigenstate pairs, $S^P_{(r_1,r_2)}=\{ (r_3,r_4) |
\omega_{1,2}=\omega_{3,4} \land R_{r_1,r_2; r_3,r_4} (m)\not =0 \}$,
whose density matrix elements $\rho_{r_3,r_4}^{B,\rm red}(m;t)$ are
coupled by Eq.\ \eqref{ODDMMasterEqu} to $\rho_{r_1,r_2}^{B,\rm
red}(m;t)$. Let us label these pairs by the indices $\alpha =
(r_1,r_2),\beta \in S^P_{(1,2)}$, where $N_{(1,2)}$ denotes the number
of elements in $S^P_{(1,2)}$.  Then, Eq.\ \eqref{ODDMMasterEqu}
reduces to subblocks of a small matrix problem
\begin{align}
\dot{\rho}^{B,\rm red}_{\alpha}(m;t) = - \sum_{\beta\in S^P_{(1,2)}} 
R_{\alpha;\beta} (m)
\rho_{\beta}^{B,\rm red}(m;t).
\label{eq:master-subblock}
\end{align}
which can be solved by exact diagonalization of the non-symmetric
complex matrix $R_{\alpha;\beta} (m)$ in terms of complex eigenvalues
$\lambda_n$ and complex right and left eigenvectors
\cite{SaadSparseLinearSystemsBook2003}.  Expanding the initial matrix
$\rho^{B,\rm red}_{\alpha}(m;t=0)$ in these eigenvectors as a sum of
the expansion coefficient $c_{\alpha,n}$ and substituting the
solutions back into Eq.\ \eqref{eq:nrg-gf}, yields
\begin{align}
\label{eq:g-broadening}
\sum_{n} \frac{c_{(r_1,r_2),n}}{\w- (E^m_{r_1}-E^m_{r_2} +\Im \lambda_n)  + i\Re \lambda_n},
\end{align}
after a Fourier transformation into frequency space. The fact that
$\rho^{B,\rm red}_{\alpha}(m;t=0)$ is a Hermitian matrix, however,
puts some constraints on theses expansion coefficients
$c_{(r_1,r_2),n}$.  %After collecting all pairs of states $(r_3,r_4)$
%Within the NRG iteration $m$, 
%for whose excitation energies $\omega_{3,4}$ that
%are equal to $\omega_{1,2}$ (typically only a very small number of
%pairs), we calculate the corresponding subblock \ac{BRT} elements
%$R_{r_1,r_2; r_3,r_4} (m)$.  
Within the NRG iteration m, we calculate those subblock BRT elements 
$R_{r_1,r_2; r_3,r_4} (m)$  for which the excitation energies $\omega_{3,4}$ are equal to $\omega_{1,2}$ (typically only a relatively small number of
pairs).
Note that the eigenvectors in general are
complex which leads to a modification of the spectral function
compared to Eq.\ \eqref{equ:T0_Lor}: It is no longer a superposition
of different Lorentzians only since the real part of $(\w-\Delta
E+i\Re \lambda)^{-1}$ also contributes via the complex expansion
coefficients $c_{(r_1,r_2),n}$.
This, however, does not modify the spectral sum rules which can be
seen by integrating the imaginary part of the spectral function,
\begin{eqnarray}
\rho_{A,B} (\w) &=& \frac{1}{2\pi} \left( G^{r}_{A,B}(\w+i\delta) - [G^{r}_{A,B}(\w+i\delta) ]^* \right),
\label{eq:49} 
\end{eqnarray}
where the contour can be closed in the upper complex plane. Since $[G^{r}_{A,B}(\w+i\delta) ]^*$ does not have poles in the upper complex plane, it does not contribute and the terms of Eq.\ \eqref{eq:g-broadening} 
yield the established sum rule \cite{PetersPruschkeAnders2006,WeichselbaumDelft2007},
\begin{eqnarray}
\int_{-\infty}^{\infty} \rho_{A,B} (\w)  &=& \expect{[A,B]_s}
\end{eqnarray}
provided the \ac{GF} was evaluated using a complete basis
set. Therefore, any potential violation of the spectral sum rule is
related to truncation errors in the implementation of the algorithm.

\subsubsection{Practical implementation of the program}

The  implementation of our approach is with respect to many aspects very
similar to the open chain approach for non-equilibrium, see Ref.\ \cite{BoekerAnders2020}.
The core of the approach remains the  NRG where its standard implementation \cite{Wilson75,BullaCostiPruschke2008} provides a Wilson chain.
For each iteration $m$, the coupling function $\Delta_{\nu,m}(z)$ of the respective $m$-th high-energy reservoir is calculated from $\Delta_{\nu,m-1}(z)$ via a continuous fraction expansion, as
layed out in detail in Sec. II C of Ref.\ \cite{BoekerAnders2020}. 
We obtain the greater reservoir correlation function, introduced in
 Eq.\ \eqref{eq:G-greater}
\begin{eqnarray}
G^{>}_{ m,\nu}(\tau) &=& -i |t'_{m \nu }|^2 \sum_k |\lambda_{m \nu, k}|^2 f(\epsilon_{m, k}) e^{i \epsilon_{m, k} \tau}
\nonumber\\
&=& - \frac{i}{\pi} \int_{- \infty}^\infty d \e \Gamma_{m,\nu}(\e) f(\e) e^{i \e \tau}
\end{eqnarray}
and the lesser \ac{GF} analogically. With Eq.\ \eqref{eq:reservoir-gfs-FT} we finally arrive at the correlation function
\begin{eqnarray}
C_{m,\nu}(\w) &=& \Gamma_{m,\nu}(\w) f(\w) \\
&& \nonumber +  \frac{i}{\pi} \int_{- \infty}^\infty d \w' \frac{\Gamma_{m,\nu}(\w') f(\w')}{\w' - \w}
\end{eqnarray}
for the $m$-th reservoir \footnote{A more detailed calculation can be found in Chap. 4 of Ref. \cite{BoekerPhD2021}}. The correlation functions of reservoirs $\tilde m \leq m$ are then used to construct the $m$-th \ac{BRT} of Eq.\ \eqref{MasterEqu}.
Since  the contributions from reservoir with a different index 
 $\tilde m$ are independent, they can be
calculated in parallel for constructing the \ac{BRT}.
Before proceeding to the next \ac{NRG} iteration, the $m$-th \ac{BRT} is stored as a sparse matrix 
for later usage - hard drive or RAM depending on the platform.

At the moment when the \ac{NRG} has  finished the last iteration,
a number of $N-m_\text{min}$ different \acp{BRT} of comparable sizes has been collected. 
Now the backwards-run of the program is initialized at the last iteration $m=N$
in the same way as was used for calculating the sum-rule conserving NRG spectral
function \cite{PetersPruschkeAnders2006,WeichselbaumDelft2007}.
The $m$-th \ac{BRT} is 
retrieved and ordered into sparse sub-blocks $(\alpha,\beta)$
as outlined in Eq.\ \eqref{eq:master-subblock}, before diagonalizing each block-matrix independently. 
This diagonalization can be performed by an exact eigen decomposition,
since the dimension of the largest block, i.e. for $\omega_{1,2} =
\omega_{3,4} = 0$, is equal to the total number of states present at
the respective iteration. Note, however, that this largest block is
not even included in the master equation \eqref{eq:master-subblock}
for fermionic \acp{GF}. The eigenvalues $\lambda_n$ of the respective
sub-blocks, as well as the complex right and left eigenvectors, are
then used to calculate all contributions to the resulting spectral
function, as shown in Eq.\ \eqref{eq:g-broadening}. Here a
parallelization of theprogram is possible for each sub-block of the \ac{BRT}.

To partially compensate for \ac{NRG} discretization artefacts we
combine our approach with an averaging procedure called z-averaging
introduced by Yoshida et al \cite{YoshidaWithakerOliveira1990} for
spectral properties extended later to non-equilibrium dynamics
\cite{AndersSchiller2005,AndersSchiller2006}.  The basic idea is to
construct a $z$-dependent Wilson chain in modifying the first
discretisation interval from $[\Lambda^{-1},1]$ to $[\Lambda^{-z},1]$
, perform independent NRG runs and average over these results
\cite{BullaCostiPruschke2008}.  Since all runs are independent, the
z-averaging can be very easily parallelized over different nodes of
the an HPC cluster.

\section{Results}

\subsection{Definition of the SIAM}

The \ac{SIAM} is one of the paradigm models suitable for the application of
the \ac{NRG} \cite{KrishWilWilson80a,KrishWilWilson80b}. The physics
of this model is essentially understood and the thermodynamics is
accessible to the Bethe ansatz approach
\cite{SCHLOTTMANN1989,AndreiFuruyaLowenstein83}, at least in the wide
band limit. This model is defined by the Hamiltonian
\begin{eqnarray}
\label{eqn:SIAM}
H&=& \sum_\sigma  \e_{d\sigma} n_\sigma + Un_\uparrow n_\downarrow 
+ \sum_{\sigma,\k} \e_{\k\sigma} c^\dagger_{\k \sigma} c_{\k \sigma} 
\nonumber \\
&& + \sum_{\sigma,\k} ( V_{\k} d^\dagger_\sigma   c_{\k \sigma}  +  V_{\k}^* c_{\k \sigma}^\dagger d_\sigma),
\end{eqnarray}
where $n_\sigma =d^\dagger_\sigma d_\sigma^{ }$,
$d^\dagger_\sigma(d_\sigma^{ })$ creates (annihilates) a local
electron with spin $\sigma$ and the single particle energy $ \e_{d
\sigma}= \e_d -\sigma H$ in the $d$-orbital. $H$ denotes the external
magnetic field, and $U$ the local Coulomb repulsion in the
$d$-orbital. This orbital hybridizes with a non-interacting conduction
band with a dispersion $\e_{\k\sigma}$ via the hybridization matrix
element $V_{\k} $. Note that we included the electron $g$-factor and
the Bohr magneton into the definition of $H$ that is measured in units
of energy. The influence of the bath coupling onto the local dynamics
is fully determined by the hybridization function
\begin{eqnarray}
\Delta_\sigma(z) &=& \sum_{\k} \frac{| V_{\k}|^2}{z- \e_{\k\sigma} },
\end{eqnarray}
where $\Gamma=\Im \Delta_\sigma(-i\delta)$ defines the charge fluctuation
scale in the problem. Applying the equation of motion, the \ac{GF}
takes the exact analytical form \cite{BullaHewsonPruschke98}
\begin{eqnarray}
\label{eq:EOM-GF}
G_{d_\sigma,d^\dagger_\sigma}(z)&=& \frac{1}{z-  \e_{d\sigma} - \Delta_\sigma(z)  -\Sigma^{U}_\sigma(z)},
\end{eqnarray}
where the correlation self-energy $\Sigma^{U}_\sigma(z)$ can be expressed by the ratio
\begin{eqnarray}
\label{eq:Sigma-U}
\Sigma^{U}_\sigma(z)&=&  U \frac{G_{d_\sigma n_{-\sigma},d^\dagger_\sigma}(z)}{G_{d_\sigma,d^\dagger_\sigma}(z)} 
.
\end{eqnarray}
The real part $\Re \Sigma^{U}_\sigma(\w-i\delta)$ contains the Hartree
part $U\langle n_{-\sigma}\rangle$ while $\Im
\Sigma^{U}_\sigma(\w-i\delta)\propto (c\pi T^2 +\w^2)$ accounts for
the local Fermi liquid properties
\cite{Yamada1974,yamadaSIAMPerturbIV75}.

\begin{figure}[t]
\begin{center}
\includegraphics[width=0.5\textwidth]{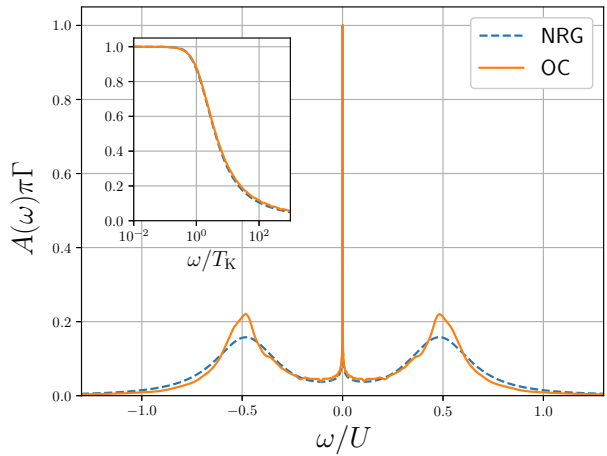}

\caption{Spectral function $A(\w)$ in dimensionless units $\w/U$ for a
particle-hole symmetric \ac{SIAM} with $U/\Gamma=20$, $D/\Gamma=100$ for
$T/T_{\rm K}=5.75 \cdot 10^{–3}$ and a Kondo temperature of $T_{\rm K}/\Gamma = 5.50 \cdot 10^{–4}$. The \ac{NRG} parameters are
$\Lambda=2$, $N_s=1000$ and $N = 50$. A z-averaging with $N_z = 12$
has been performed. The solid line represents our hybrid open chain
(OC) approach without artificial broadening, the dashed line shows the
conventional \ac{NRG} result with a Gaussian broadening parameter
$b=0.5$ - see the literature
\cite{PetersPruschkeAnders2006,WeichselbaumDelft2007,BullaCostiPruschke2008}
for details.  The inset shows the Kondo resonance.  }

\label{fig:CCOC}
\end{center}
\end{figure}

\subsection{Spectral function}

\subsubsection{Results for the spectral function of the symmetric \ac{SIAM}}

We present a comparison of results obtained by the standard \ac{NRG}
approach
\cite{PetersPruschkeAnders2006,WeichselbaumDelft2007,BullaCostiPruschke2008}
with the usual artificial broadening (blue, dashed line) and our
\ac{BR} based approach which does not require such additional
assumptions (orange solid line). The solid line has been obtained
after replacing the expression
\begin{eqnarray}
\rho^{B,\rm red}_{s,r}(m,t=0)
&=& \sum_{s'}\left[ B_{s,s'}(m) \rho^{\rm red}_{s',r}(m) \right.
\\
&& \phantom{ \sum_{s'}}
- \left. s \rho_{s,s'}^{\rm red} (m)B_{s',r}(m)\right] 
\nonumber
\end{eqnarray}
on the right hand side of Eq.\ \eqref{eq:nrg-gf} by its time dependent
\ac{BR} solution $\rho^{B,\rm red}_{s,r}(m,t)$ and analytically
performing the Fourier transformation.  Note that no artificial
broadening was needed.

Figure \ref{fig:CCOC} shows the different spectral functions
\begin{eqnarray} A_\sigma(\w)=
\rho_{d_\sigma,d^\dagger_\sigma}(\omega) = - \frac{1}{\pi} \Im
G_{d_\sigma,d^\dagger_\sigma}(\w-i\delta)
\end{eqnarray} obtained for a particle-hole symmetric \ac{SIAM} with
$U/\Gamma=20$ and a featureless conduction band, $\Im
\Delta(\w-i\delta) = \Gamma \Theta( D-|\w|)$, where $D/\Gamma=100$.
Since we consider no external magnetic field, i.\ e.\ $H=0$, we define
$A(\omega) = A_\sigma(\omega)$ for both spins $\sigma \in \{\uparrow,
\downarrow\}$. The effective temperature of the system is $T/T_{\rm
K}=5.75 \cdot 10^{–3}$ with a Kondo temperature of $T_{\rm K}/\Gamma =
5.50 \cdot 10^{–4}$. Throughout the paper, we used Wilson's definition
\cite{Wilson75} of $T_{\rm K}$, i.~e.~$\mu_{\rm eff}^2(T_{\rm
K})=0.07$, where the effective moment $\mu_{\rm eff}^2(T) = T
\chi_{\rm imp}(T)$ is linked to the impurity spin susceptibility
$\chi_{\rm imp}(T)$ \cite{BullaCostiPruschke2008}.

Since even the standard \ac{NRG} raw spectral function depends on the
artificial broadening parameter $b$ (see review
\cite{BullaCostiPruschke2008} for the technical details), the \ac{NRG}
spectral function typically presented in the literature is obtained
from the Dyson equation  Eq.\ \eqref{eq:EOM-GF}. 
In this equation, the  self-energy correction $\Sigma^{U}_\sigma(z)$ stated in Eq.\ \eqref{eq:Sigma-U}
enters. The effect of the broadening partially cancels
\cite{BullaHewsonPruschke98}. For this
reason we calculated the required NRG \acp{GF} to obtain  $\Sigma^{U}_\sigma(z)$ in all
approaches and use this $\Sigma^{U}_\sigma(z)$ 
in   Eq.\ \eqref{eq:EOM-GF} for obtaining the final spectrum.

\begin{figure}[t]
\begin{center}
\includegraphics[width=0.43\textwidth]{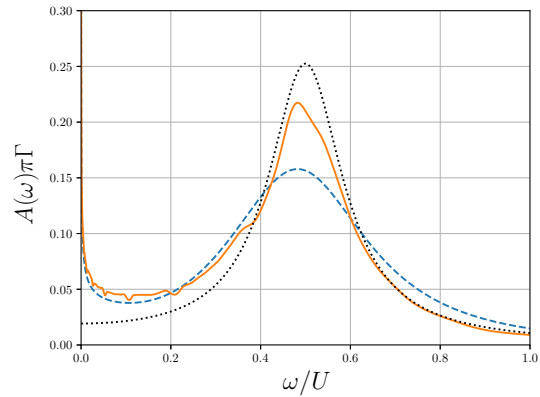}

\caption{Spectral function $A(\w)$ in dimensionless units $\w /
U$. Parameters and color coding are identical to
Fig. \ref{fig:CCOC}. The solid lines represent our hybrid OC approach 
without artificial broadening, the dashed lines show the
conventional \ac{NRG} result with a Gaussian broadening parameter
$b=0.5$. The black dashed line is a fit for the $U/\Gamma \to \infty$
Hubbard-peak at $\w = U/2$.  }

\label{fig:Hubbard}
\end{center}
\end{figure}

Figure \ref{fig:Hubbard} focuses on the Hubbard-peak for $\w>0$ and
depicts the same data as in Fig.\ \ref{fig:CCOC}. In the large $U$
limit, an analytic calculation predicts a Lorentzian shape with a
width of $2 \Gamma$, added to the plot as a black dashed curve.  For
an calculation of the well known width $2 \Gamma$ of the SIAM
see Appendix \ref{sec:SIAM-atomic-limit}. The standard \ac{NRG}
approach (dashed blue line) generates a well-known over-broadening
\cite{GreweSchmittJabbenAnders2008}. While the raw \ac{NRG} spectrum
is even broader (not shown), the application of Eq.\ \eqref{eq:EOM-GF}
leads to a narrowing that is still significantly too wide at high
energies. The \ac{BRA} (orange solid line) further reduces the
over-broadening of the Hubbard-peak. 
We would like to point out that the structures of the Hubbard peaks
also visible in Fig.\ \ref{fig:CCOC} are an artefact of the approach and dependents
on $N_s$, $\Lambda$ and also on the number of $z$-values included in the averaging procedure.
The width of the Hubbard peaks obtained with the standard approach \cite{PetersPruschkeAnders2006,WeichselbaumDelft2007} also converges to $2\Gamma$ but requires a very large number of $N_z\gg 100$ combined with an adequately adjusted artificial broadening parameter $b$. This procedure was used \cite{JovchevAnders2013} to reveal the
sharp electron-phonon peaks in the Holstein model at large electron phonon coupling where
we choose $N_z=512$ and $b=0.03$.

By reducing the discretization
parameter $\Lambda \to 1^+$ the curve further converges to the
analytic prediction (not shown).

\subsubsection{Impact of the z-averaging}

\begin{figure}[t]
\begin{center}
\includegraphics[width=0.43\textwidth]{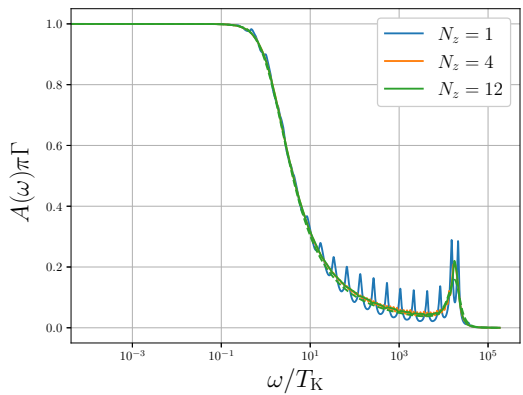}

\caption{Spectral function $A(\w)$ in dimensionless units
$\w/T_\text{K}$ with a logarithmic horizontal axis. Parameters are
identical to Fig. \ref{fig:CCOC}. The solid lines represent our hybrid
approach without artificial broadening, the dashed lines show the
conventional \ac{NRG} result with a Gaussian broadening parameter
$b=0.5$. The effect of z-averaging is shown by variation of the number
$N_z$ of configurations.  }

\label{fig:Nz}
\end{center}
\end{figure}

The data in Fig.\ \ref{fig:CCOC} and \ \ref{fig:Hubbard}, was obtained
by averaging over $N_z=12$ different $z$ values
\cite{YoshidaWithakerOliveira1990,AndersSchiller2005,AndersSchiller2006}
in the \ac{NRG}.  The necessity of z-averaging is illustrated in Fig.\
\ref{fig:Nz}. Since the \ac{BRA} only includes the reservoir coupling
in second order (see Sec. \ref{sec:Bloch-Redfield-GFs}.) the line
width of the NRG exciations acquired from the \ac{BRT} turns out to be
too small. This effectively results in unphysical oscillations of the
spectral function curves as depicted in the blue curve for $N_z = 1$,
i.~e.~in the absence of z-averaging. Varying the $z$-values in the
Wilson chain between $0<z\le 1$ continuously shifts the excitation
spectrum of the conduction band tight-binding model.  Averaging over
several \ac{NRG} runs for different z-values
\cite{YoshidaWithakerOliveira1990} can  smooth these
oscillations ($N_z > 1$ curves of Fig.\ \ref{fig:Nz}).  Here the
\ac{BR} curves basically align with the artificially broadened
ones. For $N_z = 12$ this smoothing effect has basically reached its
limit and no significant improvement can be observed by increasing
$N_z$.  To further compensate for the weak coupling of the reservoirs
in second order, one would need to decrease $\Lambda$, as discussed
above. An extension of the \ac{BRA} to fourth order coupling is
another option to improve the spectrum, yet  it does not appear to be
feasible from our current point of view - at least if no additional
symmetries, such as a diagonal \ac{BRT}, are considered.

Although the \ac{BR} equations of the composite operator needed for
the calculation of the spectral functions are independent of each
other for different $z$-values, and we only include matrix elements
within each shell, the computational effort to determine $R_{r_1,r_2;
r_3,r_4} (m)$ scales roughly as $O(m(4N_s)^4)$: The larger the
iteration index $m$ the more additional reservoirs have to be included
into the tensor elements when evaluating Eqs.\ \eqref{ODDMMasterEqu}.
The smaller $\Lambda$is,  the more \ac{NRG} states $N_s$ must be kept
after each iteration \cite{BullaCostiPruschke2008} to justify the
discarding of the high energy states. On the other hand, the spectral
weight of the auxiliary reservoir couplings is reduced when reducing
$\Lambda$ \cite{BoekerAnders2020}. This implies that for smaller
$\Lambda$ more \ac{NRG} states contribute to the spectrum with higher
energy resolution.  Due to the scaling of the \ac{BRT} calculation
with $N_s$, this would become numerically extremely expensive so 
we present a compromise between a practical CPU run-time and a
reasonable accuracy of the calculation to prove the usefulness of the
approach leaving a better optimized implementation of the algorithm to
the future.

We restrict  ourselves to the case of low $T \leq 0.01 \Gamma$ and, therefore,
considerably long Wilson chains of $N=50$ as explained in Sec.\
\ref{subsec:BRA}, and we can thus assume the off-diagonal elements of the
\ac{BRT} to be irrelevant. This improves the efficiency of the method,
since we can apply Eq.\ \eqref{eq:BRT_diag} in this case \footnote{We
have analytically shown in Appendix \ref{sec:SIAM-atomic-limit} how
the off-diagonal \ac{BRT} elements are needed in the case of the
atomic limit, i.\ e.\ $N=0$, to generate the correct line width for
$U=0$.}.  To put the gain in speed into perspective, we have
calculated the spectral function for the non-interacting case and
without z-averaging (not shown). We treat energies as degenerate if
$|E^m_{r} - E^m_s| \leq 10^{-15} \Gamma$.  In this case, the
deviations of the spectral functions between the approaches with and
without off-diagonal elements of the \ac{BRT} are below $10^{-6}
\Gamma^{-1}$ within the interval $|\w| < 10^{-5} \Gamma$ for a
temperature $T \approx 10^{-6} \Gamma$. The diagonal approximation
reduced program runtime by a factor of $6$.

 \subsubsection{Finite external magnetic field}

\begin{figure}[htbp]
\begin{center}
(a)  \includegraphics[width=0.43\textwidth]{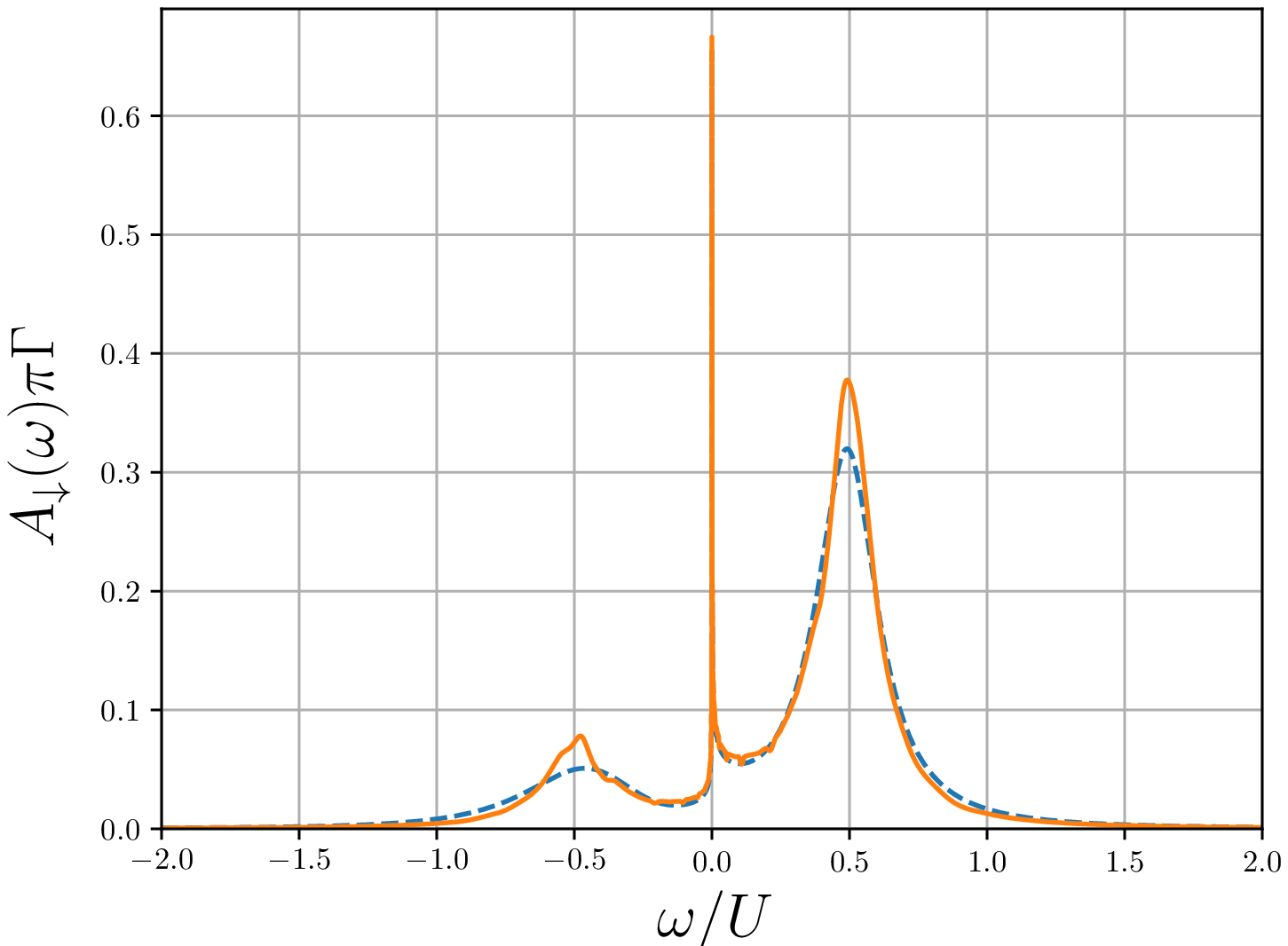}

(b)  \includegraphics[width=0.43\textwidth]{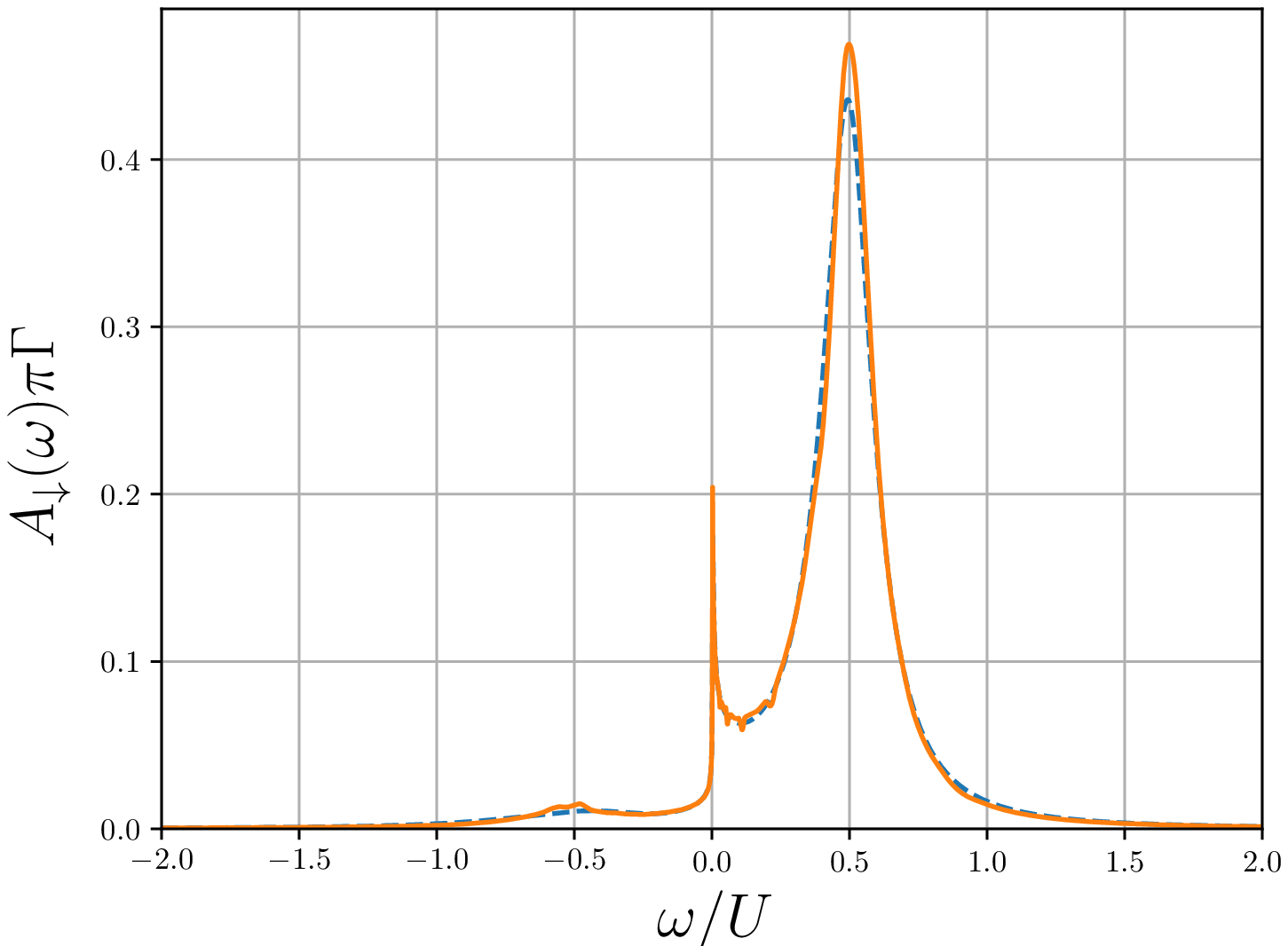}

\caption{Spectral function $A(\w)$ of the minority spin in
dimensionless units vs $\w/U$ a in a magnetic field of (a)
$H/T_\text{K}=5$ and (b) $H/T_\text{K}=100$, respectively. Parameters
and color coding as Fig.~\ref{fig:CCOC}.  }
\label{fig:H-5-100}
\end{center}
\end{figure}

\begin{figure}[t]
\begin{center}
(a) \includegraphics[width=0.43\textwidth]{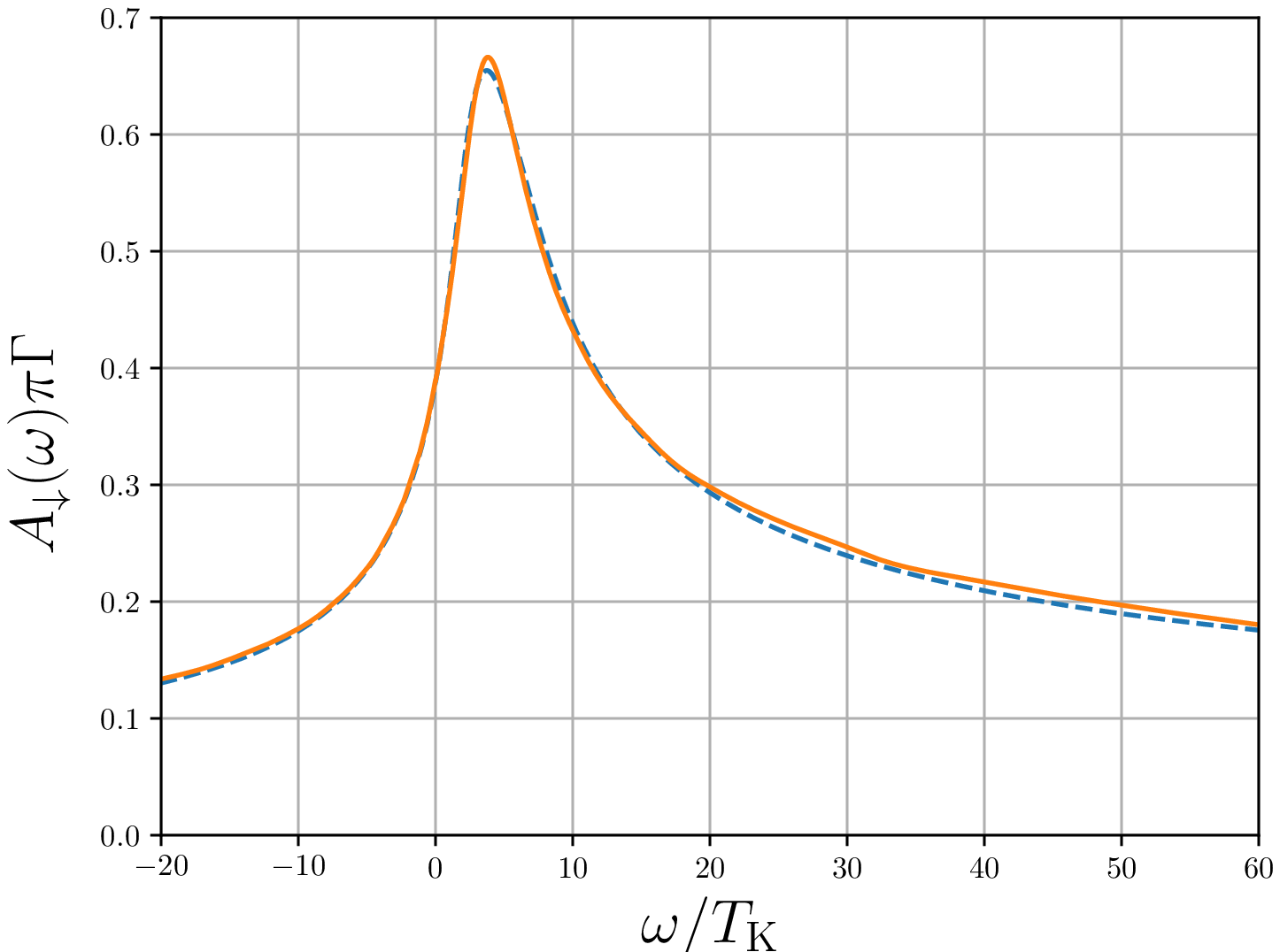}

(b) \includegraphics[width=0.43\textwidth]{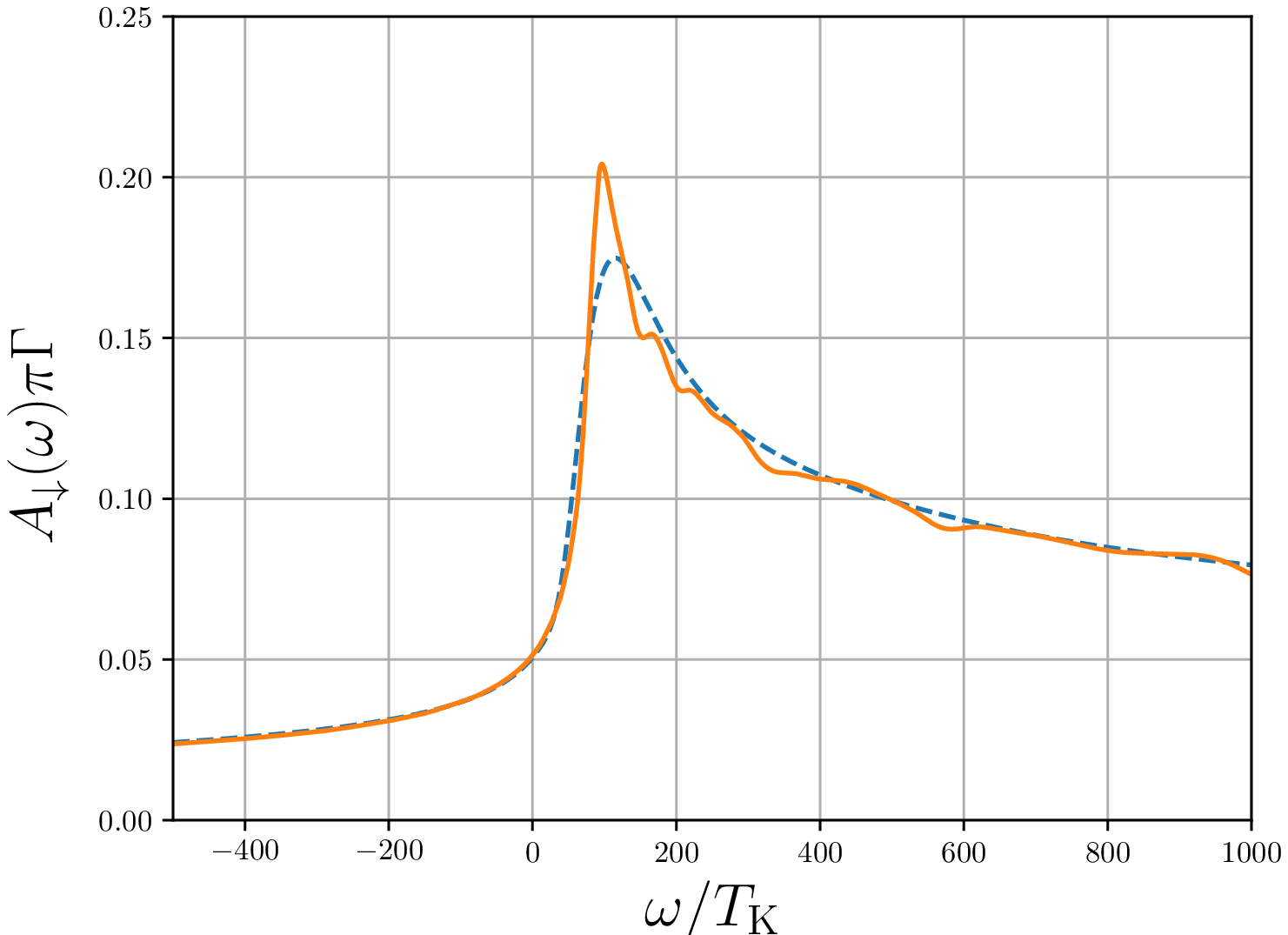}
\caption{Data of Fig \ref{fig:H-5-100}  plotted vs $\w/T_\text{K}$,
(a) $H/T_\text{K}=5$ and (b) $H/T_\text{K}=100$.
Parameters and color coding as Fig.~\ref{fig:CCOC}.
}

\label{fig:H-5-100-zoom}
\end{center}
\end{figure}

In Fig.\ \ref{fig:H-5-100} we present the minority spin spectral
function versus  $\w/U$ at a finite magnetic field $H/T_\text{K}=5$ and
$H/T_\text{K}=100$, respectively, for the model parameters of Fig.\
\ref{fig:CCOC}. Note that for a particle-hole symmetric model
considered here $A_\uparrow(\w)=A_\downarrow(-\w)$ holds. Therefore,
it is sufficient to show only one of the two spectra since they are
symmetry related.
All other parameters and color coding are identical to those in Fig.\
\ref{fig:CCOC}.  A reminiscence of the zero-frequency Kondo resonance
with a reduced peak height is still visible.  In addition, spectral
weight is shifted from the Hubbard peak at $\w\approx -U/2$ to
$\w\approx U/2$ \cite{Hofstetter2000}. For $H/T_\text{K}=100$, the
lower Hubbard peak is almost completely absent. Regardless of the
magnetic field strength, the peaks are sharper in the open chain case.

Using the same data as depicted in Fig.\ \ref{fig:H-5-100},
Fig.~\ref{fig:H-5-100-zoom} focuses on the details of the evolution of
the Kondo-resonance in a finite magnetic field. The minority spin
resonance is shifted to higher energies. At moderate fields, the
difference between the standard NRG approach and our open-chain
\ac{BRA} is insignificant. However, within large fields,
$H/T_\text{K}=100$, the asymmetry becomes more pronounced and the
\ac{BR} spectrum is much sharper around the excitation threshold of
$\w\approx H$. \cite{roschSpectralKM03,Schoeller2009a}.

\section{Conclusion}

We presented an approach to spectral functions of \acp{QIS} that
eliminates the requirement for an artificial broadening of the Lehmann
representation due to the discretization of the \ac{QIS}.  The
starting point was the exact representation of the original bath
coupling function by a Wilson chain augmented with a set of
high-energy reservoirs \cite{SBMopenchain2017,BoekerAnders2020}.  By
adapting the \ac{BRA} \cite{MayKuehn2000,JeskeBRT2015} to \acp{GF}
based on the recently presented open chain TD-NRG approach
\cite{BoekerAnders2020} we are able to include life-time effects on
the \ac{NRG} excitations induced by bath coupling which is neglected
in the pure Wilson chain.

Since the approach is based on the complete basis for the Wilson chain
\cite{AndersSchiller2005,AndersSchiller2006}, the spectral sum-rules
are always exactly fulfilled, independent of the number $N_s$ of kept
states after each \ac{NRG} iteration and independent of the new decay
channels introduced by the additional reservoirs.

By combining our \ac{BRA} to \acp{GF} with z-averaging and the exact
relations obtained from the equation of motion
\cite{BullaHewsonPruschke98} we obtain very accurate spectral
functions from the \ac{NRG} without further broadening parameters. We
gauged the quality of our spectra for $T\to 0$ with the results of a
standard \ac{NRG} approach \cite{PetersPruschkeAnders2006}.  Our
approach tracks the low energy Kondo-resonance very accurately and
produces high energy Hubbard peaks that are much narrower that the
standard \ac{NRG} spectrum: Our artificial broadening free open-chain
spectra approach the analytic prediction of a Lorentzian of width 
$2\Gamma$ \cite{Grewe83} for large $U$.

We also presented spectra in a finite magnetic field. While for small
fields, i.\ e.\ $H\approx T_\text{K}$, the open chain approach
essentially agrees with the standard \ac{NRG} results, significant
deviations are observed for $H/T_\text{K}> 50$: the spectral
properties become much sharper and much more pronounced around
$\w\approx \pm H$ and slowly approach the analytic predictions
\cite{roschSpectralKM03} for the Kondo model in very large fields.

Although our approach is tailored towards the partitioning of the NRG
eigenbasis, the general strategy can readily be adapted to other
discretized representations of \acp{QIS} such as the \ac{ED} and the
D-DMRG \cite{White92,Schollwoeck-2005,Schollwoeck2011}. For the D-DMRG
the hybridization function is typically treated by a continuous
fraction expansion \cite{RaasUhrigAnders04} producing a finite
tight-binding chain augmented by a single reservoir at the end of the
chain that is omitted in the standard method. In contrast to the
Wilson chain, for this tight-binding case the dynamics of the
composite operator $\hat \rho^{I}_{B,{S}}(t)$ as stated in
Eq.~\eqref{eq:dgl-rho-red-B-S} still hold, but the reservoir coupling
functions vanish with the exception of the last chain site $\tilde
m=N$. In ED and in the D-DMRG one might proceed in constructing a
suitable Krylov space by applying the Hamiltonian onto a starting
vector suitable for the \ac{GF} of interest and calculate the
dynamics of the operator $\rho^{I}_{B,{S}}(t)$ in this reduced Krylov
space \cite{NoceraKrylovDMRG2016} to include life-time effects of the
reservoirs neglected in the conventional approaches.

\section{Acknowledgments}
We acknowledge fruitful discussions with Jan von Delft and Saurabh
Pradhan, and financial support by Deutsche Forschungsgemeinschaft via
the Grant No.\ AN 275/10-1.

\appendix

\section{Simple example: Spectral function of the SIAM}

\label{sec:SIAM-atomic-limit}

Let us consider the minimal Hamiltonian for $H_{S}$ in the \ac{SIAM},
\begin{eqnarray} H_{S}&=&H_\text{imp} = \sum_\sigma \e_{d \sigma}
d^\dagger_\sigma d_\sigma + U n_\uparrow n_\downarrow
\end{eqnarray} and treat the full bath continuum of the SIAM by the
\ac{BRA}.  The four eigenstates $\ket{\alpha}$ with
$\alpha=0,\sigma,2$ span a four dimensional Fock space of the
impurity.  The Hubbard operators
$X_{\alpha,\beta}=\ket{\alpha}\bra{\beta}$ mediate transitions from
the local state $\ket{\beta}$ to the state $\ket{\alpha}$. We also
omit the external magnetic field and, therefore, set $\e_{d \sigma}
=\e_{d } $.  Since the Kondo effect is caused by an entanglement
between the impurity states and the conduction band continuum, no
Kondo resonance can be found within this approximation.

For the local \ac{GF} $G_{d_\sigma,d^\dagger_\sigma}(z)$, the initial
composite operator, $\hat \rho_{B,{S}}^{I}(t=0)$, is given by
\begin{eqnarray}
\hat \rho_{B,S}^{I}(0) &=& d^\dagger_\sigma\hat \rho_{S} +  \hat \rho_{S} d^\dagger_\sigma
\nonumber \\
&=& X_{\sigma,0} (\rho_{0,0}+\rho_{\sigma,\sigma} )-\sigma X_{2,-\sigma} (\rho_{-\sigma,-\sigma}+\rho_{2,2})
\nonumber \\
&=& X_{\sigma,0} A_{\sigma,0} -\sigma X_{2,-\sigma} B_{2,-\sigma},
\end{eqnarray}
where $\rho_{i,j}$ are the matrix elements of $\hat \rho_{S}$.
This reduced density operator transforms as a local fermionic creation operator and can be used
to evaluate
\begin{align}
{\rm Tr}\left[d_\sigma \hat \rho_{B}(t)\right] &=
{\rm Tr}_{S}\left[d_\sigma e^{-iH_{S} t} 
\hat \rho_{B,{S}}^{I}(t)
e^{iH_{S} t} 
\right]
\\
&= e^{i(E_0 -E_\sigma)t} A_{\sigma,0}(t) -\sigma e^{i(E_{-\sigma}- E_{2})t} B_{2,-\sigma}(t).
\nonumber
\end{align}

The time dependent matrix elements $A_{\sigma,0}(t) = \bra{\sigma}
\hat \rho_{B,{S}}^{I}(t) \ket{0}$ and $B_{2,-\sigma}(t) = \bra{2} \hat
\rho_{B,{S}}^{I}(t) \ket{-\sigma}$ contain the information about the
operator decay caused by the coupling to the reservoirs.  Apparently,
we need to distinguish two cases: if both excitation energies, $E_0
-E_\sigma$ and $E_{-\sigma}- E_{2}$, respectively, are equal, then the
exponential prefactor can be pulled out, and we are left with a
combined dynamics of $A_{\sigma,0}(t) -\sigma B_{2,-\sigma}(t)$, whose
initial condition is given by $A_{\sigma,0}(0) -\sigma
B_{2,-\sigma}(0)=1$. This is the case for $U=0$, indicating that
different density matrix elements corresponding to the same excitation
energy are coupled by Eq.\ \eqref{ODDMMasterEqu}.

For finite $U$ both excitation energies are different. In this case,
the matrix elements decouple and we find the solutions
\begin{eqnarray}
 A_{\sigma,0}(t) &=& e^{-t R_ {\sigma,0; \sigma,0}}  A_{\sigma,0}(0)\\
  B_{2,-\sigma}(t) &=& e^{-t R_ {2,-\sigma;2,- \sigma}}   B_{2,-\sigma}(0)
\end{eqnarray}
which leaves us to calculate the \ac{BRT} elements.

We focus on the regime where $E_\sigma <E_2,E_0$, and, therefore,
local moments can develop. In this case, the correct solution would
show a Kondo resonance that must be absent in this perturbative
approach.
At low temperatures, i.\ e., $|\beta\e_d|\gg 1$ we find for a constant
hybridization function in a symmetric band of width $D\gg |\e_d|$
\begin{eqnarray}
R_ {\sigma,0; \sigma,0} &=& 2\Gamma +i2\Delta E,
\end{eqnarray}
where the Lamb shift in the presence of the continuum is given by
\begin{eqnarray}
\Delta E &=& \frac{\Gamma}{\pi} \ln\left(\frac{|D+\e_d|}{|\e_d|}\right)
\end{eqnarray}
Likewise,
\begin{eqnarray}
R_ {2,-\sigma,2,- \sigma} &=& 2\Gamma -i2\Delta \bar E
\\
\Delta \bar E &=& \frac{\Gamma}{\pi} \ln\left(\frac{|D-(\e_d+U)|}{|\e_d+U|}\right)
\end{eqnarray}
Substituting these analytic solutions back into Eq.\
\eqref{eq:nrg-gf-def-36} or even Eq.\ \eqref{eq:6}, calculating the
trace and performing the Fourier transformations leads to two
contributing decaying poles
\begin{eqnarray}
G^\text{r}_{d_\sigma,d^\dagger_\sigma}(\w+i\delta)&=&
\frac{ A_{\sigma,0}(0)} { \w+i\delta - \e_d - 2\Delta E +i2\Gamma}
\\
&&\nonumber
+
\frac{ B_{2,-\sigma}(0)} { \w+i\delta -(\e_d+U) + 2\Delta \bar E +i2\Gamma}.
\end{eqnarray}
For a particle-hole symmetric impurity we have $\Delta \bar E = \Delta
E$ since $\e_d=-(\e_d+U)$, and, therefore, both poles are shifted
symmetrically by the Lamb-shift $\Delta E$. The fractional spectral
weights are also symmetric and add up to $1$.  Note that the width of
the excitations is given by $2\Gamma$ which agrees with the analytical
prediction by Grewe \cite{Grewe83}.

For $U=0$ one has to be more careful when evaluating the rate equation
\eqref{ODDMMasterEqu}. Since the two excitation energies
$E_{\sigma,0}= E_{2,-\sigma}=\e_{d}$ are equal, the two matrix
elements $\rho^B_{\sigma,0}$ and $\rho^B_{2,-\sigma}$ are coupled. The
differential equation for $A_{\sigma,0}(t) -\sigma B_{2,-\sigma}(t)$
is rather simple and is derived from the dynamics of $A_{\sigma,0}(t)
$ and $B_{2,-\sigma}(t)$.  For a constant density of states, it is
straight forward to show that we recover the exact \ac{GF}
\begin{eqnarray}
G^\text{r}_{d_\sigma,d^\dagger_\sigma}(\w+i\delta)&=&\frac{1}{ \w+i\delta - \e_d +\delta E +i \Gamma}
\end{eqnarray}
where
\begin{eqnarray}
\delta E= \frac{\Gamma}{\pi} \ln\left|\frac{D-\e_\text{d}}{D+\e_\text{d}}\right|
\end{eqnarray}
is the known contribution from the real part of the self-energy.  Note
that the width of the GF is reduced to $\Gamma$, compared to $2\Gamma$
in the case of a finite $U$.  The reduction originates in the negative
sign of the term including $B_{2,-\sigma}(t)$, which enters the
calculation of the \acp{GF}. In the combined differential equations
there is a compensation of different terms contributing to the
individual \ac{BRT} elements defined in Eq.\ \eqref{MasterEqu}.

This leaves the question what causes the discontinuity between the
solutions for $U=0$ and for $U\not =0$.  This discontinuity is related
to the secular approximation. When $U\ll \Gamma$, the approximation
does not hold since the excitation energies are small compared to the
inverse time scale of the decay of the correlation functions.  In this
case, we have to resort to a full time dependence in Eq.\
\eqref{eq:dgl-rho-red-B-S}.  In the particle-hole symmetric case, one
can show \cite{BoekerPhD2021} that this leads to the
integro-differential equation
\begin{eqnarray}
\dot   A_{\sigma,0}(t) &=& -\frac{2 \Gamma}{\pi} \int_0^t ds \frac{\sin(Ds)}{s} e^{-i U s/2}
\\
&& \times \left[
2 A_{\sigma,0}(t)  -  e^{-iU(t-s)} A^*_{\sigma,0}(t)
\right]
\nonumber
\end{eqnarray}
where $ A_{\sigma,0}(t)=B^*_{2,-\sigma}(t)$.

Extending the local system to a small but finite Wilson chain removes
the discontinuity at $U=0$ in the spectral functions obtained in the
secular approximation. For more details see Ref.\
\cite{BoekerPhD2021}.

%% -----------------------------------------------------------------
%%-----------------------------------------------------------------

%%\bibliography{hybrid-references}

%%\bibliography{hybridNrgRefs}
%\bibliography{references}

%apsrev4-2.bst 2019-01-14 (MD) hand-edited version of apsrev4-1.bst
%Control: key (0)
%Control: author (8) initials jnrlst
%Control: editor formatted (1) identically to author
%Control: production of article title (0) allowed
%Control: page (0) single
%Control: year (1) truncated
%Control: production of eprint (0) enabled
%

\end{document}